\documentclass[%
 reprint,
 amsmath,amssymb,
 aps,
]{revtex4-2}

\usepackage{graphicx}
\usepackage{dcolumn}
\usepackage{bm}

\begin{document}

\preprint{APS/123-QED}

\title{Nonequilibrium relaxation inequality on short timescales}


\author{Andrea Auconi}

\affiliation{%
 Ca’ Foscari University of Venice, DSMN - via Torino 155, 30172 Mestre (Venice), Italy
}%

\date{\today}

\begin{abstract}
An integral relation is derived from the Fokker-Planck equation which connects the steady-state probability currents with the dynamics of relaxation on short timescales in the limit of small perturbation fields. As a consequence of this integral relation, a lower bound on the steady-state entropy production is obtained. For the particular case of an ensemble of random perturbation fields of weak spatial gradient, a simpler bound is derived from the integral relation which provides a feasible method to estimate entropy production from relaxation experiments.
\end{abstract}



\maketitle


The Fokker-Planck equation describes the dynamics of the probability distribution of a Brownian particle in a force field \cite{risken1996fokker}. 
Thermodynamic nonequilibrium in the Fokker-Planck equation is defined by the presence of probability currents, which may result from nonconservative driving forces, time-dependent potentials, or in the relaxation to a steady state  \cite{ito2024geometric,dechant2022geometric,pigolotti2017generic,van2010three}.  
A central measure of nonequilibrium is dissipation or irreversible entropy production, which is linked to the statistical irreversibility of trajectories by fluctuation theorems \cite{jarzynski2011equalities,sagawa2012nonequilibrium,parrondo2015thermodynamics,seifert2019stochastic}.

The thermodynamic uncertainty relation establishes that in a nonequilibrium steady-state the entropy production can be estimated from the statistics of fluctuating time-integrated currents  \cite{barato2015thermodynamic,dechant2018current,horowitz2020thermodynamic,falasco2020unifying, falasco2020dissipation,lucente2022inference,dieball2023direct}.
However, the experimental measurement of fluctuations is often not feasible and a quest for alternative methods based on macroscopic observables has been raised \cite{freitas2022emergent}.
One such macroscopic observable could be the probability distribution itself if considered as a normalized density of identical non-interacting particles.

Signatures of nonequilibrium have been found in the dynamics of relaxation to the steady-state density,
and in particular nonequilibrium was shown to reduce the slowest timescale \cite{coghi2021role,duncan2017using,rey2015irreversible,bao2023universal},
and to impact all timescales through the spectrum of the dynamics generator in the master equation \cite{kolchinsky2024thermodynamic}.
Characterizing the full relaxation spectrum is however challenging in a continuous setting like the Fokker-Planck equation where the number of timescales involved can diverge.



In this Letter, the relaxation properties of the Fokker-Planck equation are studied in the small perturbation regime but for the shorter timescales where the dynamics is more macroscopic. The impact of steady-state currents on relaxation is here characterized with an integral relation, and from it a lower bound on the steady-state entropy production is derived. A tractable ensemble of random perturbation fields of weak spatial gradient is then proposed for which a simpler bound for the estimation of entropy production is derived.

\paragraph*{Fokker-Planck equation.}

Let us consider the position vector $\boldsymbol{x}\in \mathbb{R}^d$ of an overdamped Brownian particle \cite{risken1996fokker} which evolves according to the stochastic differential equation
\begin{equation}\label{Langevin}
    d\boldsymbol{x} = \mathbf{F} dt + \sqrt{2T} d\mathbf{W},
\end{equation}
where $\mathbf{F}\equiv \mathbf{F}(\boldsymbol{x})$ is a force vector field independent of time, $T$ is the temperature (scaled by the Boltzmann constant) which is here taken constant in time and space, and $d\mathbf{W}$ denotes vector Brownian motion increments \cite{karatzas2012brownian}. The mobility is set to unity and not considered.

\begin{figure*}
    \centering
    \includegraphics[trim={1cm 0 2cm 0},clip,scale=0.65]{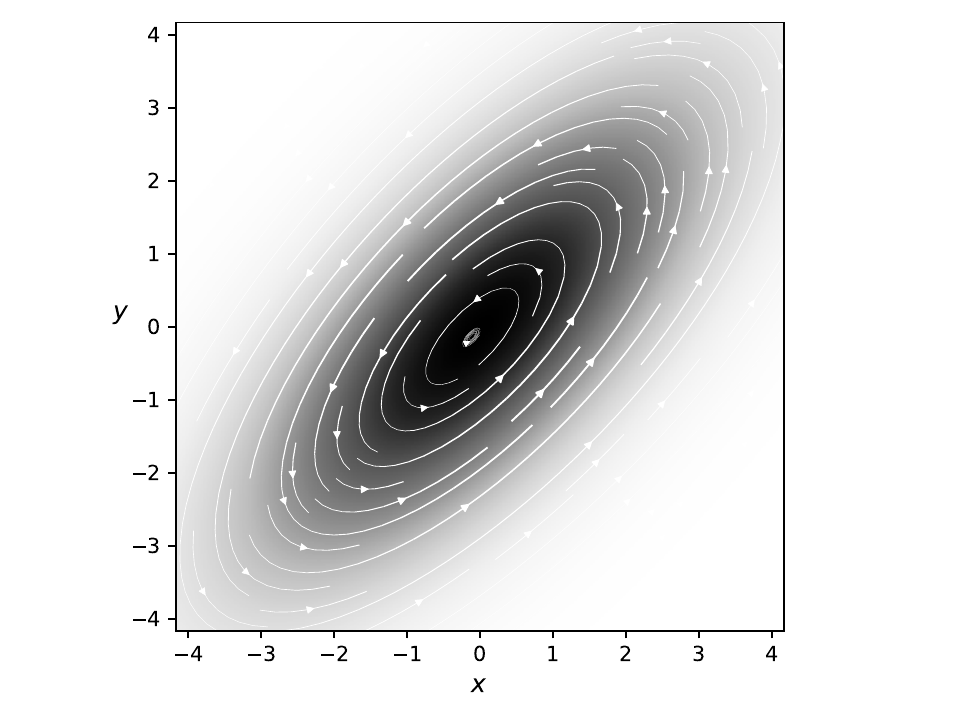} 
    \includegraphics[trim={1cm 0 2cm 0},clip,scale=0.65]{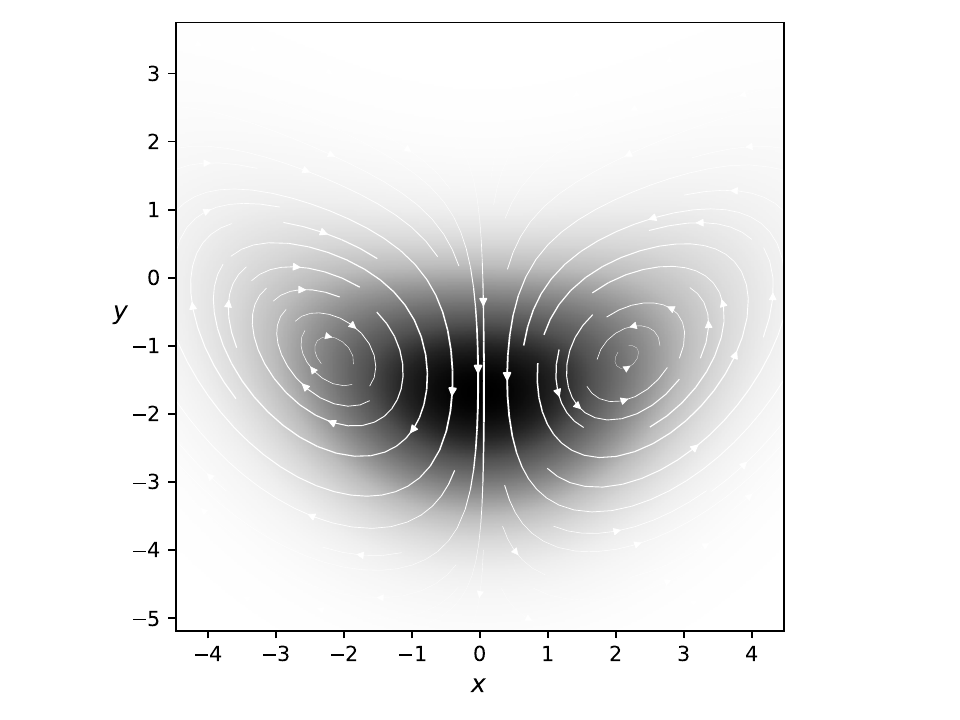} 
    \caption{\textbf{Nonequilibrium steady-states.} Prototypical examples of NESS, where arrows represent probability currents $p^*\boldsymbol{\nu}^*$ and the background color the density $p^*$.
    (Left panel) Basic signal-response model given by the force field $\boldsymbol{F}=[-\beta x, \gamma (x-y)]$, with $\beta = 1$, $\gamma = 3$, and $T=5$. Here currents are perpendicular to probability gradients, $\boldsymbol{\nu}^*\cdot \boldsymbol{\nabla}p^*=0$, while in general this holds only at the expectation level, $\langle\boldsymbol{\nu}^*\cdot \boldsymbol{\nabla}\ln p^*\rangle = 0$.
    (Right panel) Nonlinear example within the class of signal-response models, $\boldsymbol{F}=[-\beta x,\, \gamma (\mu_1 x^2 - \mu_2 -y)]$,
    with $\mu_1=2/15$ and $\mu_2 = 2$. More general examples in SM.}
    \label{fig:1}
\end{figure*}

Consider the probability density $p\equiv p(\boldsymbol{x},t)$ of a particle following the stochastic dynamics of Eq. \eqref{Langevin}. This density evolves according to the Fokker-Planck equation \cite{risken1996fokker,ito2024geometric},
\begin{equation}\label{FP1}
    \partial_t p = -\boldsymbol{\nabla}\cdot \left( p\boldsymbol{\nu} \right),
\end{equation}
that is a continuity equation for the probability current $p\boldsymbol{\nu}$ written in terms of the local mean velocity 
\begin{equation}\label{FP2}
    \boldsymbol{\nu} = \mathbf{F} -T \boldsymbol{\nabla}\ln p,
    \end{equation}
which is composed of respectively drift and diffusion. Let us assume the drift $\mathbf{F}$ to be such that a steady-state distribution $p^*$ exists, $\partial_t p^* \equiv \partial_t p|_{p^*}= 0$.
If the distribution is at steady-state $p^*$ and currents are still nonzero, $\boldsymbol{\nu}^*\neq 0$, then the system is said to be in a nonequilibrium steady-state (NESS) \cite{risken1996fokker}, see Fig. (\ref{fig:1}) for graphical examples.
The symbol $^*$ denotes quantities evaluated at steady state and therefore time-constants.



For every NESS there exists a corresponding equilibrium steady-state having the same density $p^*$ but zero currents \cite{hatano2001steady,dechant2020fluctuation,dechant2021continuous}. This is obtained by changing the force to
\begin{equation}\label{equilibrium force}
    \mathbf{F}^{eq} = T \boldsymbol{\nabla}\ln p^*,
\end{equation}
corresponding to the conservative potential $-T \ln p^*$. Note that to determine the equilibrium force $\mathbf{F}^{eq}$ the observation of currents is not required. 

The irreversible entropy production rate $\sigma \equiv \sigma (t)$ for the Fokker-Planck equation \eqref{FP1}-\eqref{FP2} is defined as
\begin{equation}
    \sigma = \frac{1}{T}\int d\boldsymbol{x} \, p \, || \boldsymbol{\nu}||^2 ,
\end{equation}
and it is interpreted as the total entropy variation rate in the system and its coupled heat bath \cite{ito2024geometric}.

\paragraph*{Linear regime.}
Consider a smooth probability perturbation field $\phi\equiv \phi (\boldsymbol{x},t)$ around the steady state,
\begin{equation}
    p = p^* (1+\phi),
\end{equation}
and assume it to be small, $|\phi|<<1$, meaning a Radon–Nikodym derivative close to $1$ everywhere. Probability normalization implies $\langle \phi \rangle\equiv \int d\boldsymbol{x}\, p^* \phi =0$, where brackets denote expectations with respect to the steady-state distribution.

By expanding Eqs. \eqref{FP1}-\eqref{FP2} in the small perturbation limit $|\phi|\rightarrow 0$ one obtains
\begin{equation}\label{phi dynamics}
    \partial_t \phi = T \nabla^2 \phi - (\boldsymbol{\nu}^* -T\boldsymbol{\nabla} \ln p^*) \cdot \boldsymbol{\nabla} \phi,
\end{equation}
that is the perturbation field dynamics close to the steady-state. Note that $\boldsymbol{\nu}^* -T\boldsymbol{\nabla} \ln p^*$ is a vector field evaluated at steady-state and therefore constant in time.

If we could control the initial state of the perturbation field $\phi(0)$ and precisely measure its instantaneous relaxation dynamics $\partial_t\phi$, then Eq. \eqref{phi dynamics} suggests a simple method to infer NESS currents $p^*\boldsymbol{\nu}^*$.
Indeed, taking a set of $d$ perturbations $(\phi_i)_{i = 1, ... ,d}$ with linearly independent gradients $(\boldsymbol{\nabla} \phi_i (\boldsymbol{x}))_{i = 1, ... ,d}$ in a point $\boldsymbol{x}$ makes Eq. \eqref{phi dynamics} into an algebraic system to determine the $d$ components of $\boldsymbol{\nu}^*(\boldsymbol{x})$.
If the same set of just $d$ perturbations has linearly independent gradients for almost every point $\boldsymbol{x}\in \mathbb{R}^d$, then it is sufficient to reconstruct the full NESS mean local velocity field $\boldsymbol{\nu}^*$ and determine the corresponding entropy production $\sigma^*$.



\paragraph*{Stability.}

Consider the Kullback-Leibler divergence \cite{dembo1991information} from the steady-state $D \equiv D[\phi]$ as a coarse-grained measure of the perturbation field,
\begin{equation}
    D \equiv  \int d\boldsymbol{x} \, p  \ln ( {p}/{p^*} ) = \frac{1}{2}\left\langle \phi^2 \right\rangle,
\end{equation}
where the last expression is valid to leading perturbation order.
The time evolution of the divergence $d_t D \equiv dD/dt$ is calculated as
\begin{equation}\label{stability}
    d_t D = \left\langle \phi \partial_t \phi \right\rangle = -T\left\langle || \boldsymbol{\nabla} \phi ||^2 \right\rangle \leq 0,
\end{equation}
where integration by parts was performed assuming $p^*\phi \boldsymbol{\nabla} \phi$ and $p^*\phi^2 \boldsymbol{\nu}^*$ decay fast enough at infinity.
We see that $d_t D$ is nonpositive, and from the normalization $\langle \phi \rangle=0$ is clear that it attains zero only when $\phi=0$ everywhere, which ensures the stability of the nonequilibrium steady-state for small perturbations. In other words, Eq. \eqref{stability} means that for the dynamics and assumptions made the Glansdorff-Prigogine criterion for stability is satisfied \cite{ito2022information,maes2015revisiting,glansdorff1974thermodynamic}.
Note that $d_t D$ is independent of the NESS currents $\boldsymbol{\nu}^*$.

\paragraph*{Nonequilibrium relaxation.}

The analysis on the shorter timescales is continued by considering the second time derivative of the divergence, which is calculated as
\begin{equation}\label{second_derivative}
    d_t^2 D = 2 T \left\langle \alpha^2 T  - \alpha \boldsymbol{\nu}^* \cdot \boldsymbol{\nabla} \phi  \right\rangle,
\end{equation}
where $\alpha \equiv \boldsymbol{\nabla} \ln p^*\cdot \boldsymbol{\nabla} \phi +\nabla^2 \phi $  is an integrand independent of the NESS currents $\boldsymbol{\nu}^*$. 
Let us denote by $D^{eq}$ the divergence of the corresponding equilibrium dynamics obtained from Eq. \eqref{equilibrium force} and starting from the same initial perturbation $\phi(0)$.
The nonequilibrium impact on the second derivative is then written
\begin{equation}\label{xi}
    \xi\equiv d_t^2 D -d_t^2 D^{eq}
    = -2T \left\langle \alpha \boldsymbol{\nu}^* \cdot \boldsymbol{\nabla} \phi  \right\rangle ,
\end{equation}
which is an integral relation connecting the relaxation dynamics on short timescales to the probability currents at steady-state through their projections on perturbation gradients, and it is the first main result of this Letter.

Taking the square of Eq. \eqref{xi}, applying the absolute value majorization and the Cauchy-Schwarz inequality first to the dot product and then to the integral, a lower bound on the steady-state entropy production is derived, 
\begin{equation}\label{single_realization_inequality}
   \sigma^*\geq \frac{\xi^2}{4T^3  \left\langle \alpha^2 || \boldsymbol{\nabla} \phi ||^2  \right\rangle}  ,
\end{equation}
valid for relaxation dynamics from small perturbation fields. The squared nonequilibrium correction to the second derivative $\xi^2$ is a property of relaxation on short timescales, and the integral $\left\langle \alpha^2 || \boldsymbol{\nabla} \phi ||^2  \right\rangle$ is a property of the applied initial perturbation field.

\paragraph*{Random fields with weak spatial gradients in 2D.}
In the following, starting back from the integral relation of Eq. \eqref{xi}, a particular ensemble of random perturbation fields is considered for its analytical tractability, and a simpler thermodynamic bound is derived.  


Let us consider periodic perturbation fields of the form
\begin{equation}\label{periodic form}
    \phi(0) = \epsilon \sin (\boldsymbol{k}\cdot \boldsymbol{x} + \varphi) +\eta,
\end{equation}
where $\eta\equiv \eta(\epsilon,\boldsymbol{k},\varphi)$ ensures probability normalization, $\langle \phi \rangle =0$. The wave vector is written $\boldsymbol{k}= k \boldsymbol{e^{i\theta}}$, where $\boldsymbol{e^{i\theta}}=(\cos(\theta),\sin(\theta))$ is the direction unit vector in two dimensions, and $k>0$ is the spatial frequency.
Note that this functional form is imposed for the initial state $\phi(t=0)$, while the dynamics could modify it for $t>0$.
Please also note that the linear regime required above for deriving Eq. \eqref{xi} here translates to having a small perturbation strength $\epsilon \ll 1$.

To leading order in $k$,  meaning for spatially slowly varying perturbation fields where statistically $k\ll ||\boldsymbol{\nabla} \ln p^*||$, the nonequilibrium relaxation correction of Eq. \eqref{xi} becomes
\begin{equation}\label{xi approx}
    \xi = -2T \epsilon^2 k^2 \cos^2 (\varphi)  \left\langle (\boldsymbol{e^{i\theta}}\cdot \boldsymbol{\nu}^*) (\boldsymbol{e^{i\theta}}\cdot \boldsymbol{\nabla} \ln p^*) \right\rangle ,
\end{equation}
so that the phase $\varphi$ effectively controls the gradient strength in the relevant region where the density is distributed.



Let us consider many replicas of the relaxation experiment with fixed perturbation strength $\epsilon$ and spatial frequency $k$, but each time with different phase $\varphi$ and direction $\theta$. 
Assuming both angles to be uniformly distributed one obtains $\mathbb{E}\xi =0$, where integration by parts was performed assuming that $p^* (\ln p^*) \boldsymbol{\nu}^*$ decays fast enough at infinity.

To highlight the nonequilibrium impact on relaxation consider the variance of the correction, $\mathrm{Var}\left[\xi\right]\equiv\mathbb{E}\xi^2 -(\mathbb{E}\xi)^2$, which is calculated to leading order in $k$ as
\begin{widetext}
\begin{equation}\label{E xi2}
    \mathrm{Var}(\xi)  = \frac{3}{16} T^2\epsilon^4 k^4  \int\int d\boldsymbol{x} d\boldsymbol{y} \left[(\boldsymbol{\nu}^*_{\boldsymbol{x}} \cdot \boldsymbol{\nu}^* _{\boldsymbol{y}}) (\boldsymbol{\nabla}p^*_{\boldsymbol{x}} \cdot \boldsymbol{\nabla}p^* _{\boldsymbol{y}}) +(\boldsymbol{\nu}^*_{\boldsymbol{x}} \cdot \boldsymbol{\nabla}p^* _{\boldsymbol{y}}) (\boldsymbol{\nabla}p^*_{\boldsymbol{x}} \cdot \boldsymbol{\nu}^* _{\boldsymbol{y}})\right],
\end{equation}
\end{widetext}
where the below trigonometric integral was used,
\begin{multline}\label{trig}
    \frac{4}{\pi} \int_0^{2\pi}d\theta \, (\mathbf{a}\cdot \boldsymbol{e^{i\theta}} )(\mathbf{b}\cdot \boldsymbol{e^{i\theta}} ) (\mathbf{c}\cdot \boldsymbol{e^{i\theta}} ) (\mathbf{d}\cdot \boldsymbol{e^{i\theta}}) \\
    =  (\mathbf{a}\cdot \mathbf{b}) ( \mathbf{c}\cdot \mathbf{d} )
    +(\mathbf{a}\cdot \mathbf{c}) ( \mathbf{b}\cdot \mathbf{d} )
    +(\mathbf{a}\cdot \mathbf{d}) ( \mathbf{c}\cdot \mathbf{b} ) ,
\end{multline}
and one can now appreciate that the simple periodic form of the perturbation field chosen in Eq. \eqref{periodic form} was useful for the analytical evaluation of spatial correlations.
Higher order terms neglected in Eqs. \eqref{xi approx}-\eqref{E xi2} are listed in the Supplementary Materials (SM), as well as an alternative expression for $\mathrm{Var}(\xi)$. 

\begin{figure}
    \centering
    \includegraphics[trim={0.3cm 0 0 0},clip,scale=0.53]{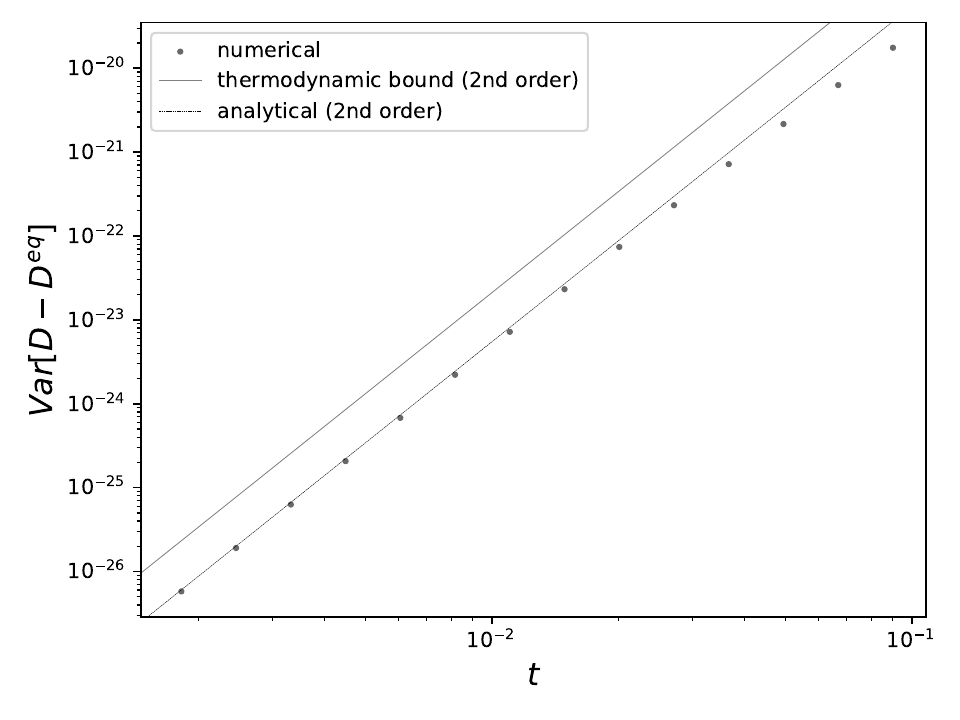}    \caption{\textbf{Linear example.} Variance of the nonequilibrium impact on relaxation in the basic signal-response model  with $\beta = 1$, $\gamma = 3$, $T=5$, and perturbations parameters $\epsilon = 0.01$ and $k=0.01$.}
    \label{fig:2}
\end{figure}

\paragraph*{Thermodynamic bound.}
Let us define the spatial Fisher information \cite{dembo1991information} as
\begin{equation}
    J \equiv T^2 \left\langle ||\boldsymbol{\nabla}\ln p^*||^2 \right\rangle\geq 0,
\end{equation}
and note that it is a property of the steady-state independent of currents. Physically it is the average squared equilibrium force $J = \left\langle ||\boldsymbol{F}^{eq}||^2 \right\rangle$ needed to produce the corresponding equilibrium steady-state.

From Eq. \eqref{E xi2}, applying the absolute value majorization and the Cauchy-Schwarz inequality first to the dot products and then to the resulting squared integral, one obtains
\begin{equation}\label{thermodynamic bound}
    \sigma^* \geq \frac{8}{3}\frac{\mathrm{Var}(\xi)}{T\epsilon^4 k^4  J} ,
\end{equation}
that is a refinement of the second law of thermodynamics for the NESS entropy production based on properties of relaxation on the shorter timescales, and it is the second main result of this Letter. 
The idea is that $\epsilon^4 k^4$ is a controllable property of the ensemble of perturbations, while the spatial Fisher information $J$ can precisely be estimated from the steady-state density.

The motivation for introducing expectations instead of optimizing the lower bound of Eq. \eqref{single_realization_inequality} over the replicas is that in the presence of measurement noise the estimation of a maximum is a statistically more involved problem. Please also note that by applying the random fields expectations directly on Eq. \eqref{single_realization_inequality} one would get a less tight inequality than Eq. \eqref{thermodynamic bound}, consistent with the fact that majorization and expectation do not commute.

\paragraph*{Measuring $\xi$.}
The applicability of Eq. \eqref{thermodynamic bound} for the estimation of the NESS entropy production is based on the measurement of the nonequilibrium relaxation correction $\xi$. While it is assumed that the relaxation dynamics of the nonequilibrium system under study is observable at least at the coarse-grained level of the divergence $d_t^2 D$, the corresponding equilibrium dynamics $d_t^2 D^{eq}$ has to be predicted from the steady-state density $p^*$. For the periodic perturbation fields with weak gradient considered here, this is calculated to leading order in $k$ as
\begin{equation}
    d_t^2 D^{eq} = 2T^2 \epsilon^2 k^2 \cos^2 (\varphi) \left\langle (\boldsymbol{e^{i\theta}}\cdot \boldsymbol{\nabla}\ln p^*)^2\right\rangle .
\end{equation}

\paragraph*{Basic linear signal-response model.}
As a first example application, let us consider the stochastic dynamics of two variables $(x,y)$ where $x$ influences the dynamics of $y$ without feedback. The basic linear signal-response model \cite{auconi2017causal} is defined by the drift field $\boldsymbol{F}=[-\beta x,\, \gamma (x-y)]$, with $\beta$ and $\gamma$ positive constants. The resulting probability currents are plotted in Fig. (\ref{fig:1}) left panel. This is possibly the simplest example in the class of non-reciprocal interactions models, whose asymmetry is known to produce entropy production \cite{auconi2019information,loos2020irreversibility,zhang2023entropy}.

Numerical verification of the results of Eqs. \eqref{E xi2}-\eqref{thermodynamic bound} are shown in Fig. (\ref{fig:2}), where the points falling below the $t^2$ line are expected as both the equilibrium and nonequilibrium dynamics relax to the same steady-state, meaning $D-D^{eq}\rightarrow 0$ for $t\rightarrow \infty$ for any small perturbation.

\begin{figure}[b]
    \centering
    \includegraphics[scale=0.5]{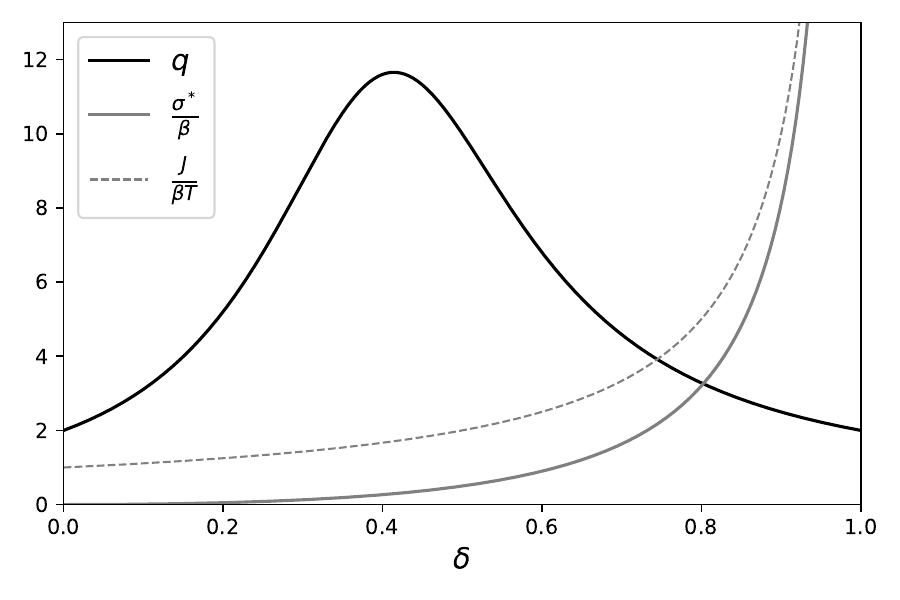}    \caption{\textbf{How tight is the bound.} 
    The basic signal-response model enables the analytical evaluation of all quantities involved, so one can immediately study how tight the thermodynamic bound here is. The tightness $q$, where $q=1$ means saturation, reaches its optimal value $q=2$ at the two domain limits, see Eq. \eqref{how tight}.}
    \label{fig:3}
\end{figure}

Analytical expressions for all the quantities involved have been derived and reported in SM. Consider here the interesting aspect of the tightness of the thermodynamic bound, that is the ratio $q$ of the entropy production with respect to the RHS of Eq. \eqref{thermodynamic bound}. For this example this is 
\begin{equation}\label{how tight}
    q = \frac{2 (1+\delta^2)^2}{4\delta^4 +(1-\delta)^4} \geq 1,
\end{equation}
where $\delta\equiv \gamma/(\gamma+\beta)$ has been defined, and accordingly $0\leq\delta \leq 1$. The tightness ratio $q$ has an optimum of $q=2$ reached at the borders of the $\delta$ domain, and it has been plotted in Fig. (\ref{fig:3}) for the reader's convenience. This $q=2$ optimum and the symmetries in the integral of Eq. \eqref{E xi2} suggest that the derivation of the thermodynamic bound could possibly be refined further.

\paragraph*{Nonlinear examples.}
A first nonlinear example is again in the class of signal-response models. This corresponds to the quadratic interaction $\boldsymbol{F}=[-\beta x,\, \gamma (\mu_1 x^2 - \mu_2 -y)]$, and the resulting probability currents are plotted in Fig. (\ref{fig:1}) right panel. The analytical formula of Eq. \eqref{E xi2} and the thermodynamic bound of Eq. \eqref{thermodynamic bound} are numerically verified and shown in SM Fig. A1.


The theoretical predictions have also been numerically verified on a more general class of random polynomial densities beyond the signal-response examples, see SM for details.

\paragraph*{Discussion.}

Signatures of nonequilibrium in the dynamics of relaxation were previously identified both in the master equation and in the Fokker-Planck equation for the longest timescale \cite{coghi2021role,duncan2017using,rey2015irreversible,bao2023universal,kolchinsky2024thermodynamic,dechant2023thermodynamic}. 
In this Letter, the entropy production in nonequilibrium steady-states of the Fokker-Planck equation has been related instead to the dynamics of relaxation on short timescales, providing a new refinement of the second law useful when fluctuations are not experimentally accessible.
The advantage of considering shorter timescales is that the dynamics is more macroscopic, while the limitation of this approach is the small perturbation regime common to linear response theory \cite{kubo2012statistical}. 

%


\bibliography{apssamp}

\providecommand{\noopsort}[1]{}\providecommand{\singleletter}[1]{#1}%
\begin{thebibliography}{36}%
\makeatletter
\providecommand \@ifxundefined [1]{%
 \@ifx{#1\undefined}
}%
\providecommand \@ifnum [1]{%
 \ifnum #1\expandafter \@firstoftwo
 \else \expandafter \@secondoftwo
 \fi
}%
\providecommand \@ifx [1]{%
 \ifx #1\expandafter \@firstoftwo
 \else \expandafter \@secondoftwo
 \fi
}%
\providecommand \natexlab [1]{#1}%
\providecommand \enquote  [1]{``#1''}%
\providecommand \bibnamefont  [1]{#1}%
\providecommand \bibfnamefont [1]{#1}%
\providecommand \citenamefont [1]{#1}%
\providecommand \href@noop [0]{\@secondoftwo}%
\providecommand \href [0]{\begingroup \@sanitize@url \@href}%
\providecommand \@href[1]{\@@startlink{#1}\@@href}%
\providecommand \@@href[1]{\endgroup#1\@@endlink}%
\providecommand \@sanitize@url [0]{\catcode `\\12\catcode `\$12\catcode `\&12\catcode `\#12\catcode `\^12\catcode `\_12\catcode `\%12\relax}%
\providecommand \@@startlink[1]{}%
\providecommand \@@endlink[0]{}%
\providecommand \url  [0]{\begingroup\@sanitize@url \@url }%
\providecommand \@url [1]{\endgroup\@href {#1}{\urlprefix }}%
\providecommand \urlprefix  [0]{URL }%
\providecommand \Eprint [0]{\href }%
\providecommand \doibase [0]{https://doi.org/}%
\providecommand \selectlanguage [0]{\@gobble}%
\providecommand \bibinfo  [0]{\@secondoftwo}%
\providecommand \bibfield  [0]{\@secondoftwo}%
\providecommand \translation [1]{[#1]}%
\providecommand \BibitemOpen [0]{}%
\providecommand \bibitemStop [0]{}%
\providecommand \bibitemNoStop [0]{.\EOS\space}%
\providecommand \EOS [0]{\spacefactor3000\relax}%
\providecommand \BibitemShut  [1]{\csname bibitem#1\endcsname}%
\let\auto@bib@innerbib\@empty
\bibitem [{\citenamefont {Risken}(1996)}]{risken1996fokker}%
  \BibitemOpen
  \bibfield  {author} {\bibinfo {author} {\bibfnamefont {H.}~\bibnamefont {Risken}},\ }\href@noop {} {\emph {\bibinfo {title} {Fokker-planck equation}}}\ (\bibinfo  {publisher} {Springer},\ \bibinfo {year} {1996})\BibitemShut {NoStop}%
\bibitem [{\citenamefont {Ito}(2024)}]{ito2024geometric}%
  \BibitemOpen
  \bibfield  {author} {\bibinfo {author} {\bibfnamefont {S.}~\bibnamefont {Ito}},\ }\bibfield  {title} {\bibinfo {title} {Geometric thermodynamics for the fokker--planck equation: stochastic thermodynamic links between information geometry and optimal transport},\ }\href@noop {} {\bibfield  {journal} {\bibinfo  {journal} {Information Geometry}\ }\textbf {\bibinfo {volume} {7}},\ \bibinfo {pages} {441} (\bibinfo {year} {2024})}\BibitemShut {NoStop}%
\bibitem [{\citenamefont {Dechant}\ \emph {et~al.}(2022)\citenamefont {Dechant}, \citenamefont {Sasa},\ and\ \citenamefont {Ito}}]{dechant2022geometric}%
  \BibitemOpen
  \bibfield  {author} {\bibinfo {author} {\bibfnamefont {A.}~\bibnamefont {Dechant}}, \bibinfo {author} {\bibfnamefont {S.-i.}\ \bibnamefont {Sasa}},\ and\ \bibinfo {author} {\bibfnamefont {S.}~\bibnamefont {Ito}},\ }\bibfield  {title} {\bibinfo {title} {Geometric decomposition of entropy production in out-of-equilibrium systems},\ }\href@noop {} {\bibfield  {journal} {\bibinfo  {journal} {Physical Review Research}\ }\textbf {\bibinfo {volume} {4}},\ \bibinfo {pages} {L012034} (\bibinfo {year} {2022})}\BibitemShut {NoStop}%
\bibitem [{\citenamefont {Pigolotti}\ \emph {et~al.}(2017)\citenamefont {Pigolotti}, \citenamefont {Neri}, \citenamefont {Rold{\'a}n},\ and\ \citenamefont {J{\"u}licher}}]{pigolotti2017generic}%
  \BibitemOpen
  \bibfield  {author} {\bibinfo {author} {\bibfnamefont {S.}~\bibnamefont {Pigolotti}}, \bibinfo {author} {\bibfnamefont {I.}~\bibnamefont {Neri}}, \bibinfo {author} {\bibfnamefont {{\'E}.}~\bibnamefont {Rold{\'a}n}},\ and\ \bibinfo {author} {\bibfnamefont {F.}~\bibnamefont {J{\"u}licher}},\ }\bibfield  {title} {\bibinfo {title} {Generic properties of stochastic entropy production},\ }\href@noop {} {\bibfield  {journal} {\bibinfo  {journal} {Physical review letters}\ }\textbf {\bibinfo {volume} {119}},\ \bibinfo {pages} {140604} (\bibinfo {year} {2017})}\BibitemShut {NoStop}%
\bibitem [{\citenamefont {Van~den Broeck}\ and\ \citenamefont {Esposito}(2010)}]{van2010three}%
  \BibitemOpen
  \bibfield  {author} {\bibinfo {author} {\bibfnamefont {C.}~\bibnamefont {Van~den Broeck}}\ and\ \bibinfo {author} {\bibfnamefont {M.}~\bibnamefont {Esposito}},\ }\bibfield  {title} {\bibinfo {title} {Three faces of the second law. ii. fokker-planck formulation},\ }\href@noop {} {\bibfield  {journal} {\bibinfo  {journal} {Physical Review E}\ }\textbf {\bibinfo {volume} {82}},\ \bibinfo {pages} {011144} (\bibinfo {year} {2010})}\BibitemShut {NoStop}%
\bibitem [{\citenamefont {Jarzynski}(2011)}]{jarzynski2011equalities}%
  \BibitemOpen
  \bibfield  {author} {\bibinfo {author} {\bibfnamefont {C.}~\bibnamefont {Jarzynski}},\ }\bibfield  {title} {\bibinfo {title} {Equalities and inequalities: Irreversibility and the second law of thermodynamics at the nanoscale},\ }\href@noop {} {\bibfield  {journal} {\bibinfo  {journal} {Annu. Rev. Condens. Matter Phys.}\ }\textbf {\bibinfo {volume} {2}},\ \bibinfo {pages} {329} (\bibinfo {year} {2011})}\BibitemShut {NoStop}%
\bibitem [{\citenamefont {Sagawa}\ and\ \citenamefont {Ueda}(2012)}]{sagawa2012nonequilibrium}%
  \BibitemOpen
  \bibfield  {author} {\bibinfo {author} {\bibfnamefont {T.}~\bibnamefont {Sagawa}}\ and\ \bibinfo {author} {\bibfnamefont {M.}~\bibnamefont {Ueda}},\ }\bibfield  {title} {\bibinfo {title} {Nonequilibrium thermodynamics of feedback control},\ }\href@noop {} {\bibfield  {journal} {\bibinfo  {journal} {Physical Review E}\ }\textbf {\bibinfo {volume} {85}},\ \bibinfo {pages} {021104} (\bibinfo {year} {2012})}\BibitemShut {NoStop}%
\bibitem [{\citenamefont {Parrondo}\ \emph {et~al.}(2015)\citenamefont {Parrondo}, \citenamefont {Horowitz},\ and\ \citenamefont {Sagawa}}]{parrondo2015thermodynamics}%
  \BibitemOpen
  \bibfield  {author} {\bibinfo {author} {\bibfnamefont {J.~M.}\ \bibnamefont {Parrondo}}, \bibinfo {author} {\bibfnamefont {J.~M.}\ \bibnamefont {Horowitz}},\ and\ \bibinfo {author} {\bibfnamefont {T.}~\bibnamefont {Sagawa}},\ }\bibfield  {title} {\bibinfo {title} {Thermodynamics of information},\ }\href@noop {} {\bibfield  {journal} {\bibinfo  {journal} {Nature physics}\ }\textbf {\bibinfo {volume} {11}},\ \bibinfo {pages} {131} (\bibinfo {year} {2015})}\BibitemShut {NoStop}%
\bibitem [{\citenamefont {Seifert}(2019)}]{seifert2019stochastic}%
  \BibitemOpen
  \bibfield  {author} {\bibinfo {author} {\bibfnamefont {U.}~\bibnamefont {Seifert}},\ }\bibfield  {title} {\bibinfo {title} {From stochastic thermodynamics to thermodynamic inference},\ }\href@noop {} {\bibfield  {journal} {\bibinfo  {journal} {Annual Review of Condensed Matter Physics}\ }\textbf {\bibinfo {volume} {10}},\ \bibinfo {pages} {171} (\bibinfo {year} {2019})}\BibitemShut {NoStop}%
\bibitem [{\citenamefont {Barato}\ and\ \citenamefont {Seifert}(2015)}]{barato2015thermodynamic}%
  \BibitemOpen
  \bibfield  {author} {\bibinfo {author} {\bibfnamefont {A.~C.}\ \bibnamefont {Barato}}\ and\ \bibinfo {author} {\bibfnamefont {U.}~\bibnamefont {Seifert}},\ }\bibfield  {title} {\bibinfo {title} {Thermodynamic uncertainty relation for biomolecular processes},\ }\href@noop {} {\bibfield  {journal} {\bibinfo  {journal} {Physical review letters}\ }\textbf {\bibinfo {volume} {114}},\ \bibinfo {pages} {158101} (\bibinfo {year} {2015})}\BibitemShut {NoStop}%
\bibitem [{\citenamefont {Dechant}\ and\ \citenamefont {Sasa}(2018)}]{dechant2018current}%
  \BibitemOpen
  \bibfield  {author} {\bibinfo {author} {\bibfnamefont {A.}~\bibnamefont {Dechant}}\ and\ \bibinfo {author} {\bibfnamefont {S.-i.}\ \bibnamefont {Sasa}},\ }\bibfield  {title} {\bibinfo {title} {Current fluctuations and transport efficiency for general langevin systems},\ }\href@noop {} {\bibfield  {journal} {\bibinfo  {journal} {Journal of Statistical Mechanics: Theory and Experiment}\ }\textbf {\bibinfo {volume} {2018}},\ \bibinfo {pages} {063209} (\bibinfo {year} {2018})}\BibitemShut {NoStop}%
\bibitem [{\citenamefont {Horowitz}\ and\ \citenamefont {Gingrich}(2020)}]{horowitz2020thermodynamic}%
  \BibitemOpen
  \bibfield  {author} {\bibinfo {author} {\bibfnamefont {J.~M.}\ \bibnamefont {Horowitz}}\ and\ \bibinfo {author} {\bibfnamefont {T.~R.}\ \bibnamefont {Gingrich}},\ }\bibfield  {title} {\bibinfo {title} {Thermodynamic uncertainty relations constrain non-equilibrium fluctuations},\ }\href@noop {} {\bibfield  {journal} {\bibinfo  {journal} {Nature Physics}\ }\textbf {\bibinfo {volume} {16}},\ \bibinfo {pages} {15} (\bibinfo {year} {2020})}\BibitemShut {NoStop}%
\bibitem [{\citenamefont {Falasco}\ \emph {et~al.}(2020)\citenamefont {Falasco}, \citenamefont {Esposito},\ and\ \citenamefont {Delvenne}}]{falasco2020unifying}%
  \BibitemOpen
  \bibfield  {author} {\bibinfo {author} {\bibfnamefont {G.}~\bibnamefont {Falasco}}, \bibinfo {author} {\bibfnamefont {M.}~\bibnamefont {Esposito}},\ and\ \bibinfo {author} {\bibfnamefont {J.-C.}\ \bibnamefont {Delvenne}},\ }\bibfield  {title} {\bibinfo {title} {Unifying thermodynamic uncertainty relations},\ }\href@noop {} {\bibfield  {journal} {\bibinfo  {journal} {New Journal of Physics}\ }\textbf {\bibinfo {volume} {22}},\ \bibinfo {pages} {053046} (\bibinfo {year} {2020})}\BibitemShut {NoStop}%
\bibitem [{\citenamefont {Falasco}\ and\ \citenamefont {Esposito}(2020)}]{falasco2020dissipation}%
  \BibitemOpen
  \bibfield  {author} {\bibinfo {author} {\bibfnamefont {G.}~\bibnamefont {Falasco}}\ and\ \bibinfo {author} {\bibfnamefont {M.}~\bibnamefont {Esposito}},\ }\bibfield  {title} {\bibinfo {title} {Dissipation-time uncertainty relation},\ }\href@noop {} {\bibfield  {journal} {\bibinfo  {journal} {Physical Review Letters}\ }\textbf {\bibinfo {volume} {125}},\ \bibinfo {pages} {120604} (\bibinfo {year} {2020})}\BibitemShut {NoStop}%
\bibitem [{\citenamefont {Lucente}\ \emph {et~al.}(2022)\citenamefont {Lucente}, \citenamefont {Baldassarri}, \citenamefont {Puglisi}, \citenamefont {Vulpiani},\ and\ \citenamefont {Viale}}]{lucente2022inference}%
  \BibitemOpen
  \bibfield  {author} {\bibinfo {author} {\bibfnamefont {D.}~\bibnamefont {Lucente}}, \bibinfo {author} {\bibfnamefont {A.}~\bibnamefont {Baldassarri}}, \bibinfo {author} {\bibfnamefont {A.}~\bibnamefont {Puglisi}}, \bibinfo {author} {\bibfnamefont {A.}~\bibnamefont {Vulpiani}},\ and\ \bibinfo {author} {\bibfnamefont {M.}~\bibnamefont {Viale}},\ }\bibfield  {title} {\bibinfo {title} {Inference of time irreversibility from incomplete information: Linear systems and its pitfalls},\ }\href@noop {} {\bibfield  {journal} {\bibinfo  {journal} {Physical Review Research}\ }\textbf {\bibinfo {volume} {4}},\ \bibinfo {pages} {043103} (\bibinfo {year} {2022})}\BibitemShut {NoStop}%
\bibitem [{\citenamefont {Dieball}\ and\ \citenamefont {Godec}(2023)}]{dieball2023direct}%
  \BibitemOpen
  \bibfield  {author} {\bibinfo {author} {\bibfnamefont {C.}~\bibnamefont {Dieball}}\ and\ \bibinfo {author} {\bibfnamefont {A.}~\bibnamefont {Godec}},\ }\bibfield  {title} {\bibinfo {title} {Direct route to thermodynamic uncertainty relations and their saturation},\ }\href@noop {} {\bibfield  {journal} {\bibinfo  {journal} {Physical Review Letters}\ }\textbf {\bibinfo {volume} {130}},\ \bibinfo {pages} {087101} (\bibinfo {year} {2023})}\BibitemShut {NoStop}%
\bibitem [{\citenamefont {Freitas}\ and\ \citenamefont {Esposito}(2022)}]{freitas2022emergent}%
  \BibitemOpen
  \bibfield  {author} {\bibinfo {author} {\bibfnamefont {J.~N.}\ \bibnamefont {Freitas}}\ and\ \bibinfo {author} {\bibfnamefont {M.}~\bibnamefont {Esposito}},\ }\bibfield  {title} {\bibinfo {title} {Emergent second law for non-equilibrium steady states},\ }\href@noop {} {\bibfield  {journal} {\bibinfo  {journal} {Nature Communications}\ }\textbf {\bibinfo {volume} {13}},\ \bibinfo {pages} {5084} (\bibinfo {year} {2022})}\BibitemShut {NoStop}%
\bibitem [{\citenamefont {Coghi}\ \emph {et~al.}(2021)\citenamefont {Coghi}, \citenamefont {Chetrite},\ and\ \citenamefont {Touchette}}]{coghi2021role}%
  \BibitemOpen
  \bibfield  {author} {\bibinfo {author} {\bibfnamefont {F.}~\bibnamefont {Coghi}}, \bibinfo {author} {\bibfnamefont {R.}~\bibnamefont {Chetrite}},\ and\ \bibinfo {author} {\bibfnamefont {H.}~\bibnamefont {Touchette}},\ }\bibfield  {title} {\bibinfo {title} {Role of current fluctuations in nonreversible samplers},\ }\href@noop {} {\bibfield  {journal} {\bibinfo  {journal} {Physical Review E}\ }\textbf {\bibinfo {volume} {103}},\ \bibinfo {pages} {062142} (\bibinfo {year} {2021})}\BibitemShut {NoStop}%
\bibitem [{\citenamefont {Duncan}\ \emph {et~al.}(2017)\citenamefont {Duncan}, \citenamefont {N{\"u}sken},\ and\ \citenamefont {Pavliotis}}]{duncan2017using}%
  \BibitemOpen
  \bibfield  {author} {\bibinfo {author} {\bibfnamefont {A.~B.}\ \bibnamefont {Duncan}}, \bibinfo {author} {\bibfnamefont {N.}~\bibnamefont {N{\"u}sken}},\ and\ \bibinfo {author} {\bibfnamefont {G.~A.}\ \bibnamefont {Pavliotis}},\ }\bibfield  {title} {\bibinfo {title} {Using perturbed underdamped langevin dynamics to efficiently sample from probability distributions},\ }\href@noop {} {\bibfield  {journal} {\bibinfo  {journal} {Journal of Statistical Physics}\ }\textbf {\bibinfo {volume} {169}},\ \bibinfo {pages} {1098} (\bibinfo {year} {2017})}\BibitemShut {NoStop}%
\bibitem [{\citenamefont {Rey-Bellet}\ and\ \citenamefont {Spiliopoulos}(2015)}]{rey2015irreversible}%
  \BibitemOpen
  \bibfield  {author} {\bibinfo {author} {\bibfnamefont {L.}~\bibnamefont {Rey-Bellet}}\ and\ \bibinfo {author} {\bibfnamefont {K.}~\bibnamefont {Spiliopoulos}},\ }\bibfield  {title} {\bibinfo {title} {Irreversible langevin samplers and variance reduction: a large deviations approach},\ }\href@noop {} {\bibfield  {journal} {\bibinfo  {journal} {Nonlinearity}\ }\textbf {\bibinfo {volume} {28}},\ \bibinfo {pages} {2081} (\bibinfo {year} {2015})}\BibitemShut {NoStop}%
\bibitem [{\citenamefont {Bao}\ and\ \citenamefont {Hou}(2023)}]{bao2023universal}%
  \BibitemOpen
  \bibfield  {author} {\bibinfo {author} {\bibfnamefont {R.}~\bibnamefont {Bao}}\ and\ \bibinfo {author} {\bibfnamefont {Z.}~\bibnamefont {Hou}},\ }\bibfield  {title} {\bibinfo {title} {Universal trade-off between irreversibility and relaxation timescale},\ }\href@noop {} {\bibfield  {journal} {\bibinfo  {journal} {arXiv e-prints}\ ,\ \bibinfo {pages} {arXiv}} (\bibinfo {year} {2023})}\BibitemShut {NoStop}%
\bibitem [{\citenamefont {Kolchinsky}\ \emph {et~al.}(2024)\citenamefont {Kolchinsky}, \citenamefont {Ohga},\ and\ \citenamefont {Ito}}]{kolchinsky2024thermodynamic}%
  \BibitemOpen
  \bibfield  {author} {\bibinfo {author} {\bibfnamefont {A.}~\bibnamefont {Kolchinsky}}, \bibinfo {author} {\bibfnamefont {N.}~\bibnamefont {Ohga}},\ and\ \bibinfo {author} {\bibfnamefont {S.}~\bibnamefont {Ito}},\ }\bibfield  {title} {\bibinfo {title} {Thermodynamic bound on spectral perturbations, with applications to oscillations and relaxation dynamics},\ }\href@noop {} {\bibfield  {journal} {\bibinfo  {journal} {Physical Review Research}\ }\textbf {\bibinfo {volume} {6}},\ \bibinfo {pages} {013082} (\bibinfo {year} {2024})}\BibitemShut {NoStop}%
\bibitem [{\citenamefont {Karatzas}\ and\ \citenamefont {Shreve}(2012)}]{karatzas2012brownian}%
  \BibitemOpen
  \bibfield  {author} {\bibinfo {author} {\bibfnamefont {I.}~\bibnamefont {Karatzas}}\ and\ \bibinfo {author} {\bibfnamefont {S.}~\bibnamefont {Shreve}},\ }\href@noop {} {\emph {\bibinfo {title} {Brownian motion and stochastic calculus}}},\ Vol.\ \bibinfo {volume} {113}\ (\bibinfo  {publisher} {Springer Science \& Business Media},\ \bibinfo {year} {2012})\BibitemShut {NoStop}%
\bibitem [{\citenamefont {Hatano}\ and\ \citenamefont {Sasa}(2001)}]{hatano2001steady}%
  \BibitemOpen
  \bibfield  {author} {\bibinfo {author} {\bibfnamefont {T.}~\bibnamefont {Hatano}}\ and\ \bibinfo {author} {\bibfnamefont {S.-i.}\ \bibnamefont {Sasa}},\ }\bibfield  {title} {\bibinfo {title} {Steady-state thermodynamics of langevin systems},\ }\href@noop {} {\bibfield  {journal} {\bibinfo  {journal} {Physical review letters}\ }\textbf {\bibinfo {volume} {86}},\ \bibinfo {pages} {3463} (\bibinfo {year} {2001})}\BibitemShut {NoStop}%
\bibitem [{\citenamefont {Dechant}\ and\ \citenamefont {Sasa}(2020)}]{dechant2020fluctuation}%
  \BibitemOpen
  \bibfield  {author} {\bibinfo {author} {\bibfnamefont {A.}~\bibnamefont {Dechant}}\ and\ \bibinfo {author} {\bibfnamefont {S.-i.}\ \bibnamefont {Sasa}},\ }\bibfield  {title} {\bibinfo {title} {Fluctuation--response inequality out of equilibrium},\ }\href@noop {} {\bibfield  {journal} {\bibinfo  {journal} {Proceedings of the National Academy of Sciences}\ }\textbf {\bibinfo {volume} {117}},\ \bibinfo {pages} {6430} (\bibinfo {year} {2020})}\BibitemShut {NoStop}%
\bibitem [{\citenamefont {Dechant}\ and\ \citenamefont {Sasa}(2021)}]{dechant2021continuous}%
  \BibitemOpen
  \bibfield  {author} {\bibinfo {author} {\bibfnamefont {A.}~\bibnamefont {Dechant}}\ and\ \bibinfo {author} {\bibfnamefont {S.-i.}\ \bibnamefont {Sasa}},\ }\bibfield  {title} {\bibinfo {title} {Continuous time reversal and equality in the thermodynamic uncertainty relation},\ }\href@noop {} {\bibfield  {journal} {\bibinfo  {journal} {Physical Review Research}\ }\textbf {\bibinfo {volume} {3}},\ \bibinfo {pages} {L042012} (\bibinfo {year} {2021})}\BibitemShut {NoStop}%
\bibitem [{\citenamefont {Dembo}\ \emph {et~al.}(1991)\citenamefont {Dembo}, \citenamefont {Cover},\ and\ \citenamefont {Thomas}}]{dembo1991information}%
  \BibitemOpen
  \bibfield  {author} {\bibinfo {author} {\bibfnamefont {A.}~\bibnamefont {Dembo}}, \bibinfo {author} {\bibfnamefont {T.~M.}\ \bibnamefont {Cover}},\ and\ \bibinfo {author} {\bibfnamefont {J.~A.}\ \bibnamefont {Thomas}},\ }\bibfield  {title} {\bibinfo {title} {Information theoretic inequalities},\ }\href@noop {} {\bibfield  {journal} {\bibinfo  {journal} {IEEE Transactions on Information theory}\ }\textbf {\bibinfo {volume} {37}},\ \bibinfo {pages} {1501} (\bibinfo {year} {1991})}\BibitemShut {NoStop}%
\bibitem [{\citenamefont {Ito}(2022)}]{ito2022information}%
  \BibitemOpen
  \bibfield  {author} {\bibinfo {author} {\bibfnamefont {S.}~\bibnamefont {Ito}},\ }\bibfield  {title} {\bibinfo {title} {Information geometry, trade-off relations, and generalized glansdorff--prigogine criterion for stability},\ }\href@noop {} {\bibfield  {journal} {\bibinfo  {journal} {Journal of Physics A: Mathematical and Theoretical}\ }\textbf {\bibinfo {volume} {55}},\ \bibinfo {pages} {054001} (\bibinfo {year} {2022})}\BibitemShut {NoStop}%
\bibitem [{\citenamefont {Maes}\ and\ \citenamefont {Neto{\v{c}}n{\`y}}(2015)}]{maes2015revisiting}%
  \BibitemOpen
  \bibfield  {author} {\bibinfo {author} {\bibfnamefont {C.}~\bibnamefont {Maes}}\ and\ \bibinfo {author} {\bibfnamefont {K.}~\bibnamefont {Neto{\v{c}}n{\`y}}},\ }\bibfield  {title} {\bibinfo {title} {Revisiting the glansdorff--prigogine criterion for stability within irreversible thermodynamics},\ }\href@noop {} {\bibfield  {journal} {\bibinfo  {journal} {Journal of Statistical Physics}\ }\textbf {\bibinfo {volume} {159}},\ \bibinfo {pages} {1286} (\bibinfo {year} {2015})}\BibitemShut {NoStop}%
\bibitem [{\citenamefont {Glansdorff}\ \emph {et~al.}(1974)\citenamefont {Glansdorff}, \citenamefont {Nicolis},\ and\ \citenamefont {Prigogine}}]{glansdorff1974thermodynamic}%
  \BibitemOpen
  \bibfield  {author} {\bibinfo {author} {\bibfnamefont {P.}~\bibnamefont {Glansdorff}}, \bibinfo {author} {\bibfnamefont {G.}~\bibnamefont {Nicolis}},\ and\ \bibinfo {author} {\bibfnamefont {I.}~\bibnamefont {Prigogine}},\ }\bibfield  {title} {\bibinfo {title} {The thermodynamic stability theory of non-equilibrium states},\ }\href@noop {} {\bibfield  {journal} {\bibinfo  {journal} {Proceedings of the National Academy of Sciences}\ }\textbf {\bibinfo {volume} {71}},\ \bibinfo {pages} {197} (\bibinfo {year} {1974})}\BibitemShut {NoStop}%
\bibitem [{\citenamefont {Auconi}\ \emph {et~al.}(2017)\citenamefont {Auconi}, \citenamefont {Giansanti},\ and\ \citenamefont {Klipp}}]{auconi2017causal}%
  \BibitemOpen
  \bibfield  {author} {\bibinfo {author} {\bibfnamefont {A.}~\bibnamefont {Auconi}}, \bibinfo {author} {\bibfnamefont {A.}~\bibnamefont {Giansanti}},\ and\ \bibinfo {author} {\bibfnamefont {E.}~\bibnamefont {Klipp}},\ }\bibfield  {title} {\bibinfo {title} {Causal influence in linear langevin networks without feedback},\ }\href@noop {} {\bibfield  {journal} {\bibinfo  {journal} {Physical Review E}\ }\textbf {\bibinfo {volume} {95}},\ \bibinfo {pages} {042315} (\bibinfo {year} {2017})}\BibitemShut {NoStop}%
\bibitem [{\citenamefont {Auconi}\ \emph {et~al.}(2019)\citenamefont {Auconi}, \citenamefont {Giansanti},\ and\ \citenamefont {Klipp}}]{auconi2019information}%
  \BibitemOpen
  \bibfield  {author} {\bibinfo {author} {\bibfnamefont {A.}~\bibnamefont {Auconi}}, \bibinfo {author} {\bibfnamefont {A.}~\bibnamefont {Giansanti}},\ and\ \bibinfo {author} {\bibfnamefont {E.}~\bibnamefont {Klipp}},\ }\bibfield  {title} {\bibinfo {title} {Information thermodynamics for time series of signal-response models},\ }\href@noop {} {\bibfield  {journal} {\bibinfo  {journal} {Entropy}\ }\textbf {\bibinfo {volume} {21}},\ \bibinfo {pages} {177} (\bibinfo {year} {2019})}\BibitemShut {NoStop}%
\bibitem [{\citenamefont {Loos}\ and\ \citenamefont {Klapp}(2020)}]{loos2020irreversibility}%
  \BibitemOpen
  \bibfield  {author} {\bibinfo {author} {\bibfnamefont {S.~A.}\ \bibnamefont {Loos}}\ and\ \bibinfo {author} {\bibfnamefont {S.~H.}\ \bibnamefont {Klapp}},\ }\bibfield  {title} {\bibinfo {title} {Irreversibility, heat and information flows induced by non-reciprocal interactions},\ }\href@noop {} {\bibfield  {journal} {\bibinfo  {journal} {New Journal of Physics}\ }\textbf {\bibinfo {volume} {22}},\ \bibinfo {pages} {123051} (\bibinfo {year} {2020})}\BibitemShut {NoStop}%
\bibitem [{\citenamefont {Zhang}\ \emph {et~al.}(2023)\citenamefont {Zhang}, \citenamefont {Garcia-Millan} \emph {et~al.}}]{zhang2023entropy}%
  \BibitemOpen
  \bibfield  {author} {\bibinfo {author} {\bibfnamefont {Z.}~\bibnamefont {Zhang}}, \bibinfo {author} {\bibfnamefont {R.}~\bibnamefont {Garcia-Millan}}, \emph {et~al.},\ }\bibfield  {title} {\bibinfo {title} {Entropy production of nonreciprocal interactions},\ }\href@noop {} {\bibfield  {journal} {\bibinfo  {journal} {Physical Review Research}\ }\textbf {\bibinfo {volume} {5}},\ \bibinfo {pages} {L022033} (\bibinfo {year} {2023})}\BibitemShut {NoStop}%
\bibitem [{\citenamefont {Dechant}\ \emph {et~al.}(2023)\citenamefont {Dechant}, \citenamefont {Garnier-Brun},\ and\ \citenamefont {Sasa}}]{dechant2023thermodynamic}%
  \BibitemOpen
  \bibfield  {author} {\bibinfo {author} {\bibfnamefont {A.}~\bibnamefont {Dechant}}, \bibinfo {author} {\bibfnamefont {J.}~\bibnamefont {Garnier-Brun}},\ and\ \bibinfo {author} {\bibfnamefont {S.-i.}\ \bibnamefont {Sasa}},\ }\bibfield  {title} {\bibinfo {title} {Thermodynamic bounds on correlation times},\ }\href@noop {} {\bibfield  {journal} {\bibinfo  {journal} {arXiv preprint arXiv:2303.13038}\ } (\bibinfo {year} {2023})}\BibitemShut {NoStop}%
\bibitem [{\citenamefont {Kubo}\ \emph {et~al.}(2012)\citenamefont {Kubo}, \citenamefont {Toda},\ and\ \citenamefont {Hashitsume}}]{kubo2012statistical}%
  \BibitemOpen
  \bibfield  {author} {\bibinfo {author} {\bibfnamefont {R.}~\bibnamefont {Kubo}}, \bibinfo {author} {\bibfnamefont {M.}~\bibnamefont {Toda}},\ and\ \bibinfo {author} {\bibfnamefont {N.}~\bibnamefont {Hashitsume}},\ }\href@noop {} {\emph {\bibinfo {title} {Statistical physics II: nonequilibrium statistical mechanics}}},\ Vol.~\bibinfo {volume} {31}\ (\bibinfo  {publisher} {Springer Science \& Business Media},\ \bibinfo {year} {2012})\BibitemShut {NoStop}%
\end{thebibliography}%


\providecommand{\noopsort}[1]{}\providecommand{\singleletter}[1]{#1}%
%

\end{document}


\preprint{APS/123-QED}

\title{Supplementary Materials for the manuscript\\ ``Nonequilibrium relaxation inequality on short timescales"}




\maketitle

\section{Derivation of Eq. (7)}

The starting point is the Fokker-Planck equation  for the overdamped Brownian particle \cite{risken1996fokker,ito2024geometric}, that is introduced in Eqs. (2)-(3) of the main text,
\begin{equation}\label{FP}
    \partial_t p = -\boldsymbol{\nabla}\cdot \left( p\boldsymbol{\nu} \right); \,\,\,\,\,\,\,\,
    \boldsymbol{\nu} = \mathbf{F} -T \boldsymbol{\nabla}\ln p.
    \end{equation}
A smooth probability perturbation field $\phi$ around the steady state has then been considered,
\begin{equation}
    p = p^* (1+\phi),
\end{equation}
and probability normalization implies $\langle \phi \rangle =0$.
In this section the dynamics of the perturbation field $\phi$ in the small perturbation limit $|\phi|\rightarrow 0$ is studied. By expanding the probability gradient one obtains
\begin{equation}
    \boldsymbol{\nabla}\ln p = \boldsymbol{\nabla}\ln p^* +\frac{\boldsymbol{\nabla} \phi}{1+\phi} = \boldsymbol{\nabla}\ln p^* +\boldsymbol{\nabla} \phi + \mathcal{O}(\phi \boldsymbol{\nabla} \phi),
\end{equation}
and then for the local mean velocity $\boldsymbol{\nu} = \boldsymbol{\nu}^* -T\boldsymbol{\nabla} \phi + \mathcal{O}(\phi \boldsymbol{\nabla} \phi)$. 
The currents divergence is expanded as
\begin{equation}\label{currents divergence expansion}
    \boldsymbol{\nabla}\cdot \left( p\boldsymbol{\nu} \right) = p^*\boldsymbol{\nu}^* \cdot \boldsymbol{\nabla} \phi -T p^* \nabla^2 \phi
-T \boldsymbol{\nabla} p^* \cdot \boldsymbol{\nabla} \phi
+ \mathcal{O}\left(\boldsymbol{\nabla}\cdot(p^* \phi \boldsymbol{\nabla} \phi)\right),
\end{equation}
where the steady-state condition property $\boldsymbol{\nabla}\cdot \left( p^*\boldsymbol{\nu}^* \right) = 0$ has been used. Note that no requirement on the gradient magnitude $||\boldsymbol{\nabla} \phi||$ is necessary here as the limit $|\phi|\rightarrow 0$ is sufficient for the terms $ \mathcal{O}\left(\boldsymbol{\nabla}\cdot(p^* \phi \boldsymbol{\nabla} \phi)\right)$ to be negligible in Eq. \eqref{currents divergence expansion}. In particular, in this limit the ratio $||\boldsymbol{\nabla}\phi||^2/\nabla^2 \phi$ vanishes as it can be shown by considering $\phi=\epsilon \widetilde \phi$ for fixed $\widetilde \phi$ in the limit $\epsilon\rightarrow 0$.
Then from $\partial_t p = p^* \partial_t \phi$ and Eqs. \eqref{FP}-\eqref{currents divergence expansion}, in the small perturbation limit $|\phi|\rightarrow 0$ one obtains
\begin{equation}\label{phi dynamics}
    \partial_t \phi = T \nabla^2 \phi - (\boldsymbol{\nu}^* -T\boldsymbol{\nabla} \ln p^*) \cdot \boldsymbol{\nabla} \phi,
\end{equation}
which is Eq. (7) in the main text.

\section{Derivation of Eq. (9)}

The Kullback-Leibler divergence from the steady-state $D \equiv D[\phi]$ was introduced as a coarse-grained measure of the perturbation field,
\begin{equation}
    D \equiv  \int d\boldsymbol{x} \, p  \ln ( {p}/{p^*} ) = \frac{1}{2}\left\langle \phi^2 \right\rangle,
\end{equation}
where the last expression is valid to leading order in the small perturbation limit $|\phi|\rightarrow 0$.
The time evolution of the divergence $d_t D \equiv dD/dt$ is written considering Eq. \eqref{phi dynamics} as
\begin{equation}
    d_t D = \left\langle \phi \partial_t \phi \right\rangle = T \left\langle \phi \nabla^2 \phi \right\rangle - \left\langle \phi \boldsymbol{\nabla} \phi \cdot (\boldsymbol{\nu}^* -T\boldsymbol{\nabla} \ln p^*) \right\rangle .
\end{equation}
Consider first the expectation
\begin{equation}
    \left\langle \phi \boldsymbol{\nabla} \phi \cdot \boldsymbol{\nabla} \ln p^* \right\rangle = \int d\boldsymbol{x} \, \phi \boldsymbol{\nabla} \phi \cdot \boldsymbol{\nabla} p^*
    = \int d\boldsymbol{x} \, \boldsymbol{\nabla}\cdot ( p^* \phi \boldsymbol{\nabla} \phi ) -  \int d\boldsymbol{x} \,p^* \boldsymbol{\nabla}\cdot ( \phi \boldsymbol{\nabla} \phi )
    = - \left\langle || \boldsymbol{\nabla} \phi ||^2 \right\rangle -\left\langle \phi \nabla^2 \phi \right\rangle,
\end{equation}
where integration by part was performed, and then the integral $\int d\boldsymbol{x} \, \boldsymbol{\nabla}\cdot ( p^* \phi \boldsymbol{\nabla} \phi )=0$ evaluated to zero by the divergence theorem with the assumption that $p^*\phi \boldsymbol{\nabla} \phi$ decays fast enough at infinity.

Let us now evaluate the integral involving the steady-state probability currents $\boldsymbol{\nu}^*$,
\begin{equation}\label{passage stability}
    \left\langle \phi \boldsymbol{\nabla} \phi \cdot \boldsymbol{\nu}^* \right\rangle
    = \int d\boldsymbol{x} \, p^* \phi \boldsymbol{\nabla} \phi \cdot \boldsymbol{\nu}^*
   = \int d\boldsymbol{x} \, \boldsymbol{\nabla} \cdot ( p^* \phi^2  \boldsymbol{\nu}^* )
   -\int d\boldsymbol{x} \, \phi \boldsymbol{\nabla} \cdot ( p^* \phi  \boldsymbol{\nu}^* ) 
   = -\left\langle \phi \boldsymbol{\nabla} \phi \cdot \boldsymbol{\nu}^* \right\rangle,
\end{equation}
where integration by part was performed assuming $p^*\phi^2 \boldsymbol{\nu}^*$ to decay fast enough at infinity, and the steady-state condition property $\boldsymbol{\nabla}\cdot \left( p^*\boldsymbol{\nu}^* \right) = 0$ was used. From Eq. \eqref{passage stability} is clear that $\left\langle \phi \boldsymbol{\nabla} \phi \cdot \boldsymbol{\nu}^* \right\rangle = 0$, and from this and the above equations one obtains
\begin{equation}\label{stability}
    d_t D = -T\left\langle || \boldsymbol{\nabla} \phi ||^2 \right\rangle,
\end{equation}
which is Eq. (9) in the main text. The nonnegativity of the square immediately implies the stability $d_t D \leq 0$.

\section{Derivation of Eq. (10)-(12)}

Let us consider the second time derivative of the divergence by differentiating Eq. \eqref{stability},
\begin{multline}\label{second_derivative}
    d_t^2 D = -2 T \left\langle \boldsymbol{\nabla} \phi \cdot \boldsymbol{\nabla}(\partial_t \phi)  \right\rangle\\
    = -2T \left[  \int d\boldsymbol{x}\, \boldsymbol{\nabla} \cdot \left[ p^* (\partial_t \phi) \boldsymbol{\nabla} \phi  \right] -\int d\boldsymbol{x}\, (\partial_t \phi) \boldsymbol{\nabla} \cdot \left( p^*  \boldsymbol{\nabla} \phi  \right)  \right]
    = 2T \left\langle \left[ \boldsymbol{\nabla} \ln p^*\cdot \boldsymbol{\nabla} \phi +\nabla^2 \phi \right] \partial_t \phi \right\rangle\\
    = 2T^2 \left\langle \left[ \boldsymbol{\nabla} \ln p^*\cdot \boldsymbol{\nabla} \phi +\nabla^2 \phi \right]^2 \right\rangle
    -2T \left\langle (\boldsymbol{\nu}^* \cdot \boldsymbol{\nabla} \phi) \left[ \boldsymbol{\nabla} \ln p^*\cdot \boldsymbol{\nabla} \phi +\nabla^2 \phi \right] \right\rangle,
\end{multline}
where integration by parts was performed assuming $p^* (\partial_t \phi) \boldsymbol{\nabla} \phi $ decays fast enough at infinity. 
Then by defining $\alpha \equiv \boldsymbol{\nabla} \ln p^*\cdot \boldsymbol{\nabla} \phi +\nabla^2 \phi $ this is rewritten as
\begin{equation}\label{second_derivative}
    d_t^2 D = 2 T \left\langle \alpha^2 T  - \alpha \boldsymbol{\nu}^* \cdot \boldsymbol{\nabla} \phi  \right\rangle,
\end{equation}
that is Eq. (10) in the main text. The nonequilibrium correction to the second derivative is then
\begin{equation}\label{xi}
    \xi\equiv d_t^2 D -d_t^2 D^{eq}
    = -2T \left\langle \alpha \boldsymbol{\nu}^* \cdot \boldsymbol{\nabla} \phi  \right\rangle ,
\end{equation}
where $D^{eq}$ denotes the divergence in the corresponding equilibrium dynamics for the same perturbation. 

The thermodynamics bound for single perturbations is then derived as
\begin{multline}
      \xi^2
    = 4T^2 \left\langle \alpha \boldsymbol{\nu}^* \cdot \boldsymbol{\nabla} \phi  \right\rangle^2 \,
    \leq 4T^2 \left\langle |\alpha|\, | \boldsymbol{\nu}^* \cdot \boldsymbol{\nabla} \phi | \right\rangle^2 \,
    \leq 4T^2 \left\langle |\alpha| \, ||\boldsymbol{\nu}^* ||\, || \boldsymbol{\nabla} \phi || \right\rangle^2 \\
    \leq 4T^2 \left\langle ||\boldsymbol{\nu}^* ||^2 \right\rangle \left\langle \alpha^2\, || \boldsymbol{\nabla} \phi ||^2 \right\rangle\,
    = 4T^3 \sigma^* \left\langle \alpha^2\, || \boldsymbol{\nabla} \phi ||^2 \right\rangle  
    ,  
\end{multline}
where the first inequality is just majorization by the absolute value, the second is the Cauchy-Schwarz inequality for the dot product, and the last is the Cauchy-Schwarz inequality for square-integrable functions.

\section{Derivation of Eq. (14) and $\mathbb{E}\xi = 0$.}

Periodic perturbations have been introduced in the main text in the form
\begin{equation}\label{periodic form}
    \phi(0) = \epsilon \sin (\boldsymbol{k}\cdot \boldsymbol{x} + \varphi) +\eta,
\end{equation}
where $\eta\equiv \eta(\epsilon,\boldsymbol{k},\varphi)$ ensures probability normalization, $\langle \phi \rangle =0$. The wave vector is written $\boldsymbol{k}= k \boldsymbol{e^{i\theta}}$, where $\boldsymbol{e^{i\theta}}=(\cos(\theta),\sin(\theta))$ is the direction unit vector, and $k>0$ is the spatial frequency.
Note that by introducing $\boldsymbol{e^{i\theta}}$ the problem has been restricted to the two-dimensional case.
Also note that this periodic functional form is imposed for the initial state $\phi(t=0)$, while the dynamics could modify it for $t>0$.

The gradient and Laplacian for this perturbation field are respectively $\boldsymbol{\nabla}\phi_0 = \epsilon \boldsymbol{k} \cos (\boldsymbol{k}\cdot \boldsymbol{x} + \varphi) $ and $\nabla^2\phi_0 = -\epsilon k^2 \sin (\boldsymbol{k}\cdot \boldsymbol{x} + \varphi)$.
The nonequilibrium relaxation correction of Eq. \eqref{xi} is then
\begin{equation}\label{xi periodic}
    \xi = -2T \epsilon^2 \left\langle (\boldsymbol{k}\cdot \boldsymbol{\nu}^*) \left[ (\boldsymbol{k}\cdot \boldsymbol{\nabla} \ln p^*) \cos^2 (\boldsymbol{k}\cdot \boldsymbol{x} + \varphi) -k^2 \cos (\boldsymbol{k}\cdot \boldsymbol{x} + \varphi) \sin (\boldsymbol{k}\cdot \boldsymbol{x} + \varphi) \right] \right\rangle.
\end{equation}
Expectations $\mathbb{E}\equiv \mathbb{E}_{\theta} \mathbb{E}_{\varphi}$ are meant with respect to the phase $\varphi$ and direction $\theta$ of the periodic perturbation field, assuming these to be uniformly distributed over $[0,2\pi)$. Given the trigonometric integrals $\int d\varphi\, \cos^2(\zeta+\varphi) = \pi$ and $\int d\varphi\, \cos(\zeta+\varphi) \sin(\zeta+\varphi) = 0$, the expectation over the phase $\varphi$ evaluates to
\begin{equation}
    \mathbb{E}_{\varphi} \xi = -T \epsilon^2 \left\langle (\boldsymbol{k}\cdot \boldsymbol{\nu}^*) (\boldsymbol{k}\cdot \boldsymbol{\nabla} \ln p^*) \right\rangle.
\end{equation}
From the same trigonometric integrals it is immediately derived that, given two arbitrary vectors $\mathbf{a}$ and $\mathbf{b}$, the below integral identity holds,
\begin{equation}
    \int_0^{2\pi} d\theta \, (\mathbf{a}\cdot \boldsymbol{e^{i\theta}}) (\mathbf{b}\cdot \boldsymbol{e^{i\theta}})
    = \pi\, \mathbf{a}\cdot\mathbf{b} ,
\end{equation}
and using it to further average $\mathbb{E}_{\varphi} \xi$ over the direction angle $\theta$ one finds
\begin{equation}
    \mathbb{E} \xi = -\frac{T \epsilon^2 k^2}{2} \left\langle \boldsymbol{\nu}^* \cdot \boldsymbol{\nabla} \ln p^* \right\rangle = 0,
\end{equation}
where the last equality is obtained via integration by parts,
\begin{equation}\label{nu_dlnp_0}
    \left\langle \boldsymbol{\nu}^* \cdot \boldsymbol{\nabla} \ln p^* \right\rangle = \int d\boldsymbol{x}\, p^* \boldsymbol{\nu}^* \cdot \boldsymbol{\nabla} \ln p^* 
    = \int d\boldsymbol{x}\, \boldsymbol{\nabla} \cdot \left( p^* (\ln p^*) \boldsymbol{\nu}^* \right) -\int d\boldsymbol{x}\, (\ln p^*) \boldsymbol{\nabla} \cdot \left( p^* \boldsymbol{\nu}^*    \right) = 0,
\end{equation}
assuming that $p^* (\ln p^*) \boldsymbol{\nu}^* $ decays fast enough at infinity.
Please note that for the results in this section no assumption on the spatial frequency $k$ was made, and indeed the requirement of spatially slowly varying perturbation fields will be needed only to obtain the analytical expression for the variance $\mathrm{Var}\left[\xi\right]$ in the next section.

\section{Derivation of Eq. (15) and the thermodynamic bound.}

Consider the square of the nonequilibrium correction of Eq. \eqref{xi periodic},
\begin{multline}
    \xi^2  = 4T^2 \epsilon^4 \int\int d\boldsymbol{x}d\boldsymbol{y}\, \,p^*_{\boldsymbol{x}}p^*_{\boldsymbol{y}} (\boldsymbol{k}\cdot \boldsymbol{\nu}^*_{\boldsymbol{x}}) (\boldsymbol{k}\cdot \boldsymbol{\nu}^*_{\boldsymbol{y}})
    \cos (\boldsymbol{k}\cdot \boldsymbol{x} + \varphi) \cos (\boldsymbol{k}\cdot \boldsymbol{y} + \varphi) \Big[ 
    (\boldsymbol{k}\cdot \boldsymbol{\nabla} \ln p^*_{\boldsymbol{x}}) (\boldsymbol{k}\cdot \boldsymbol{\nabla} \ln p^*_{\boldsymbol{y}})  \cos (\boldsymbol{k}\cdot \boldsymbol{x} + \varphi) \cos (\boldsymbol{k}\cdot \boldsymbol{y} + \varphi) \\
    -2 k^2 (\boldsymbol{k}\cdot \boldsymbol{\nabla} \ln p^*_{\boldsymbol{x}}) \cos (\boldsymbol{k}\cdot \boldsymbol{x} + \varphi) \sin (\boldsymbol{k}\cdot \boldsymbol{y} + \varphi)\,
    +k^4 \sin (\boldsymbol{k}\cdot \boldsymbol{x} + \varphi) \sin (\boldsymbol{k}\cdot \boldsymbol{y} + \varphi)\Big] .
\end{multline}
Given the below trigonometric integrals,
\begin{center}
\begin{equation}\label{trigonometric_integrals}
  \begin{gathered}
     \frac{1}{2\pi} \int_0^{2\pi} d\varphi \, \cos^2(a+\varphi) \cos^2(b+\varphi)
     = \frac{2+\cos[2(b-a)]}{8},\\
     \frac{1}{2\pi} \int_0^{2\pi} d\varphi \, \cos^2(a+\varphi) \cos(b+\varphi) \sin(b+\varphi)
     = \frac{\sin[2(b-a)]}{8},\\
     \frac{1}{2\pi} \int_0^{2\pi} d\varphi \, \cos(a+\varphi)\sin(a+\varphi)  \cos(b+\varphi) \sin(b+\varphi)
     = \frac{\cos[2(b-a)]}{8},
      \end{gathered}
\end{equation}
\end{center}
one can evaluate the expectation with respect to the phase,
\begin{multline}
    \mathbb{E}_{\varphi}\xi^2  = \frac{T^2 \epsilon^4}{2} \int d\boldsymbol{x}d\boldsymbol{y}\, \,p^*_{\boldsymbol{x}}p^*_{\boldsymbol{y}} (\boldsymbol{k}\cdot \boldsymbol{\nu}^*_{\boldsymbol{x}}) (\boldsymbol{k}\cdot \boldsymbol{\nu}^*_{\boldsymbol{y}})
 \Big[(\boldsymbol{k}\cdot \boldsymbol{\nabla} \ln p^*_{\boldsymbol{x}}) (\boldsymbol{k}\cdot \boldsymbol{\nabla} \ln p^*_{\boldsymbol{y}})  \left( 2+\cos[2\boldsymbol{k}\cdot (\boldsymbol{x}-\boldsymbol{y})] \right) \\
    +2 k^2 (\boldsymbol{k}\cdot \boldsymbol{\nabla} \ln p^*_{\boldsymbol{x}}) \sin [2\boldsymbol{k}\cdot (\boldsymbol{x} -\boldsymbol{y})] \,
    +k^4 \cos [2\boldsymbol{k}\cdot (\boldsymbol{x} -\boldsymbol{y})] \Big]\\
    = \frac{3 \,T^2 \epsilon^4}{2} \int d\boldsymbol{x}d\boldsymbol{y}\, \,p^*_{\boldsymbol{x}}p^*_{\boldsymbol{y}} (\boldsymbol{k}\cdot \boldsymbol{\nu}^*_{\boldsymbol{x}}) (\boldsymbol{k}\cdot \boldsymbol{\nu}^*_{\boldsymbol{y}}) (\boldsymbol{k}\cdot \boldsymbol{\nabla} \ln p^*_{\boldsymbol{x}}) (\boldsymbol{k}\cdot \boldsymbol{\nabla} \ln p^*_{\boldsymbol{y}})  +\mathcal{O}(k^6),
\end{multline}
where only the leading order terms in $k$ were kept, meaning that the spatially slowly varying perturbation field limit has been introduced.
The trigonometric integrals of Eq. \eqref{trigonometric_integrals} imply that, given four arbitrary vectors $\mathbf{a}$, $\mathbf{b}$, $\mathbf{c}$, and $\mathbf{d}$, the below integral identity holds,
\begin{equation}
    \frac{4}{\pi} \int_0^{2\pi}d\theta \, (\mathbf{a}\cdot \boldsymbol{e^{i\theta}} )(\mathbf{b}\cdot \boldsymbol{e^{i\theta}} ) (\mathbf{c}\cdot \boldsymbol{e^{i\theta}} ) (\mathbf{d}\cdot \boldsymbol{e^{i\theta}})
    =  (\mathbf{a}\cdot \mathbf{b}) ( \mathbf{c}\cdot \mathbf{d} )
    +(\mathbf{a}\cdot \mathbf{c}) ( \mathbf{b}\cdot \mathbf{d} )
    +(\mathbf{a}\cdot \mathbf{d}) ( \mathbf{c}\cdot \mathbf{b} ) .   
\end{equation}
From this identity, and considering $\left\langle \boldsymbol{\nu}^* \cdot \boldsymbol{\nabla} \ln p^* \right\rangle =0$ as in Eq. \eqref{nu_dlnp_0}, by further averaging $\mathbb{E}_{\varphi}\xi^2$ with respect to the direction $\theta$ one obtains Eq. (15) of the main text. From it the thermodynamic bound is obtained as follows,
\begin{multline}\label{bound derivation}
    \mathrm{Var}(\xi)  = \frac{3}{16} T^2\epsilon^4 k^4  \int\int d\boldsymbol{x} d\boldsymbol{y}\, p^*_{\boldsymbol{x}}p^*_{\boldsymbol{y}} \left[(\boldsymbol{\nu}^*_{\boldsymbol{x}} \cdot \boldsymbol{\nu}^* _{\boldsymbol{y}}) (\boldsymbol{\nabla}\ln p^*_{\boldsymbol{x}} \cdot \boldsymbol{\nabla}\ln p^* _{\boldsymbol{y}}) +(\boldsymbol{\nu}^*_{\boldsymbol{x}} \cdot \boldsymbol{\nabla}\ln p^* _{\boldsymbol{y}}) (\boldsymbol{\nabla} \ln p^*_{\boldsymbol{x}} \cdot \boldsymbol{\nu}^* _{\boldsymbol{y}})\right] \\
    \leq \frac{3}{16} T^2\epsilon^4 k^4  \int\int d\boldsymbol{x} d\boldsymbol{y}\, p^*_{\boldsymbol{x}}p^*_{\boldsymbol{y}} \left(|\boldsymbol{\nu}^*_{\boldsymbol{x}} \cdot \boldsymbol{\nu}^* _{\boldsymbol{y}}|\,| \boldsymbol{\nabla}\ln p^*_{\boldsymbol{x}} \cdot \boldsymbol{\nabla}\ln p^* _{\boldsymbol{y}} |\,+\,| \boldsymbol{\nu}^*_{\boldsymbol{x}} \cdot \boldsymbol{\nabla}\ln p^* _{\boldsymbol{y}}| \, | \boldsymbol{\nabla} \ln p^*_{\boldsymbol{x}} \cdot \boldsymbol{\nu}^* _{\boldsymbol{y}}| \right)\\
    \leq \frac{3}{8} T^2\epsilon^4 k^4  \int\int d\boldsymbol{x} d\boldsymbol{y}\, p^*_{\boldsymbol{x}}p^*_{\boldsymbol{y}} \, ||\boldsymbol{\nu}^*_{\boldsymbol{x}} || \, ||\boldsymbol{\nu}^* _{\boldsymbol{y}}||\,|| \boldsymbol{\nabla}\ln p^*_{\boldsymbol{x}} || \, || \boldsymbol{\nabla}\ln p^* _{\boldsymbol{y}} ||\\
    =\frac{3}{8} T^2\epsilon^4 k^4 \left( \int d\boldsymbol{x} \, p^* \, ||\boldsymbol{\nu}^* || \, || \boldsymbol{\nabla}\ln p^* || \right)^2 \\
    \leq \frac{3}{8} T^2\epsilon^4 k^4 \left(\int d\boldsymbol{x} \, p^* \, ||\boldsymbol{\nu}^* || ^2 \right)  \left(
    \int d\boldsymbol{x} \, p^* \, || \boldsymbol{\nabla}\ln p^* || ^2 \right)\\
    = \frac{3}{8} T\epsilon^4 k^4 \sigma^* J,
\end{multline}
where, similar to the single realizations thermodynamic bound derivation, the first inequality is just majorization by the absolute value, the second is the Cauchy-Schwarz inequality for the dot product, and the last is the Cauchy-Schwarz inequality for square-integrable functions.

\section{Expectation and majorization do not commute.}
The thermodynamic bound obtained by applying the Cauchy-Schwarz inequality after taking the expectations is tighter than that resulting from taking expectations on the bound valid for all single realizations. Consider the thermodynamic bound for single realizations, Eq. (12) in the main text,
\begin{equation}\label{single_realization_inequality}
   \sigma^*\geq \frac{\xi^2}{4T^3  \left\langle \alpha^2 || \boldsymbol{\nabla} \phi ||^2  \right\rangle}  ,
\end{equation}
and expand the integrand $\alpha^2 || \boldsymbol{\nabla} \phi ||^2$, with the periodic perturbation form considered, in the limit of small spatial frequency,
\begin{equation}
    \alpha^2 || \boldsymbol{\nabla} \phi ||^2 = \epsilon^4 k^4 \cos^4 (\varphi) \left( \boldsymbol{e^{i\theta}} \cdot \boldsymbol{\nabla}\ln p^* \right)^2
    +\mathcal{O}(k^5).
\end{equation}
Averaging $\xi^2$ and $\left\langle \alpha^2 || \boldsymbol{\nabla} \phi ||^2  \right\rangle$ separately using the first trigonometric integral of Eq. \eqref{trigonometric_integrals} and $\mathbb{E}\xi = 0$ one gets
\begin{equation}\label{less tight}
    \sigma^* \geq \frac{4\,\mathrm{Var}(\xi)}{3T\epsilon^4 k^4  J} ,
\end{equation}
which is less tight than the thermodynamic bound derived in Eq. \eqref{bound derivation}.
Note, however, that we could also take as bound the maximum over the replicas in the rhs of Eq. \eqref{single_realization_inequality}. Nevertheless, in the presence of measurement noise the estimation of a maximum is a more involved statistical problem compared to the estimation of a mean, while the focus here is on simplicity and applicability of the method.

\section{Alternative expression for $\mathrm{Var}(\xi)$.}

Define the swapped gradient as
\begin{equation}
    \boldsymbol{\nabla}_c \equiv \left[ \partial_2, \partial_1  \right],
\end{equation}
meaning the gradient $\boldsymbol{\nabla} = \left[ \partial_1, \partial_2 \right]$ with its two components swapped.
The equality of Eq. \eqref{nu_dlnp_0}, $\left\langle \boldsymbol{\nu}^* \cdot \boldsymbol{\nabla} \ln p^* \right\rangle =0$, implies for the components
\begin{equation}
    \left\langle \nu^*_1 \partial_1 \ln p^* \right\rangle = -\left\langle \nu^*_2 \partial_2 \ln p^* \right\rangle.
\end{equation}
Expanding $\mathrm{Var}(\xi)$ in the vector components combinations and rearranging terms one obtains
\begin{equation}\label{alternative}
    \mathrm{Var}(\xi)  = \frac{3}{16} T^2\epsilon^4 k^4 \left( \left\langle \boldsymbol{\nu}^* \cdot \boldsymbol{\nabla}_c \ln p^*\right\rangle^2 +4\left\langle \nu^*_1 \partial_1 \ln p^* \right\rangle^2 \right),
\end{equation}
which does not involve the double integral and its numerical implementation is more immediate.

\section{Basic linear example}
The basic linear signal-response model \cite{auconi2017causal,auconi2019information} corresponds to a drift field $\boldsymbol{F}=[-\beta x,\, \gamma (x-y)]$, where $\mathbf{x}\equiv [x,y]$ is the position vector, and $\beta$ and $\gamma$ are positive constants. 
Analytical expressions for all the quantities involved can be derived following standard techniques of the Ornstein-Uhlenbeck process \cite{risken1996fokker}, or also by using Ito's Lemma \cite{karatzas2012brownian} as it is done here.
Let us start from the coupled SDEs corresponding to this force field,
\begin{equation}
    \begin{cases}
        dx = -\beta x dt + \sqrt{2T} dW_x,\\
        dy = \gamma (x-y) dt + \sqrt{2T} dW_y,
    \end{cases}
\end{equation}
where $dW_x,dW_y$ denote uncorrelated standard Brownian motion increments \cite{karatzas2012brownian}, characterized by $\langle dW_x \rangle = 0$ and $dW_x dW_x = dt$ in the Ito notation.
All expectation values are invariant at steady-state, and this property can be used together with Ito's Lemma to derive the below expectations,
\begin{equation}
    0 = d\langle x\rangle = \langle dx\rangle = -\beta \langle x\rangle dt  \implies \langle x\rangle = 0,
\end{equation}
\begin{equation}
    0 = d\langle y\rangle = -\gamma \langle y\rangle dt \implies \langle y\rangle = 0,
\end{equation}
\begin{equation}
    0 = d\langle x^2\rangle = 2\langle x dx\rangle +\langle dx dx\rangle = -2\beta \langle x^2 \rangle dt + 2T dt \implies \langle x^2\rangle = T/\beta,
\end{equation}
\begin{equation}
    0 = d\langle x y\rangle = -(\beta+\gamma) \langle x y\rangle dt +\gamma \langle x^2 \rangle dt \implies \langle x y\rangle = (T/\beta) \delta,
    \,\,\,\,\,\,\,\,
    \delta \equiv \frac{\gamma}{\gamma+\beta},
\end{equation}
\begin{equation}
    0 = d\langle y^2 \rangle = 2\gamma \left( \langle x y\rangle -\langle y^2 \rangle \right) dt  + 2T dt \implies \langle y^2 \rangle = (T/\beta) s,
    \,\,\,\,\,\,\,\,
    s \equiv \delta+\frac{1}{\delta}-1.
\end{equation}
These are the expectations and covariance matrix entries sufficient to write the bivariate Gaussian density
\begin{equation}
    p^* = A \exp \left[ B \left( s x^2 + y^2  -2\delta x y \right) \right],
\end{equation}
with $B\equiv -\frac{\beta}{2T(s-\delta^2)}$, and $A$ normalization. The log gradient is then
\begin{equation}
    \boldsymbol{\nabla}\ln p^* = 2B \left[sx-\delta y, \, -\delta x +y  \right],
\end{equation}
and the local mean velocity is
\begin{equation}
    \boldsymbol{\nu}^* = \mathbf{F} -T \boldsymbol{\nabla}\ln p^* = \frac{\beta\delta}{s-\delta^2}  \left[ \delta x -y, \, s x-\delta y  \right] .
\end{equation}
Evaluating the terms in Eq. \eqref{alternative} with the derived expressions for the basic linear signal-response model one obtains
\begin{equation}\label{linear_E_Var_xi}
    \mathrm{Var}(\xi)  = \frac{3}{16} T^2\epsilon^4 k^4 \frac{\beta^2}{(s-\delta^2)^2} \left[ 4\delta^4 + (1-\delta)^4 \right],
\end{equation}
\begin{equation}\label{linear_sigma_J}
    \sigma^* = \frac{\beta \delta^2}{1-\delta}, \,\,\,\,\,\,\,\,  J = \frac{T\beta}{1-\delta},
\end{equation}
From Eqs. \eqref{linear_E_Var_xi}-\eqref{linear_sigma_J} we can evaluate how tight is the thermodynamic bound in this example, 
\begin{equation}
    q \equiv \frac{3T\epsilon^4 k^4 \sigma^* J}{8\mathrm{Var}(\xi)} = \frac{2 (1+\delta^2)^2}{4\delta^4 +(1-\delta)^4} \geq 1.
\end{equation}

\section{Supplementary plot: quadratic interaction.}
\begin{figure}
    \centering
    \includegraphics[trim={0.4cm 0 0 0},clip,scale=0.55]{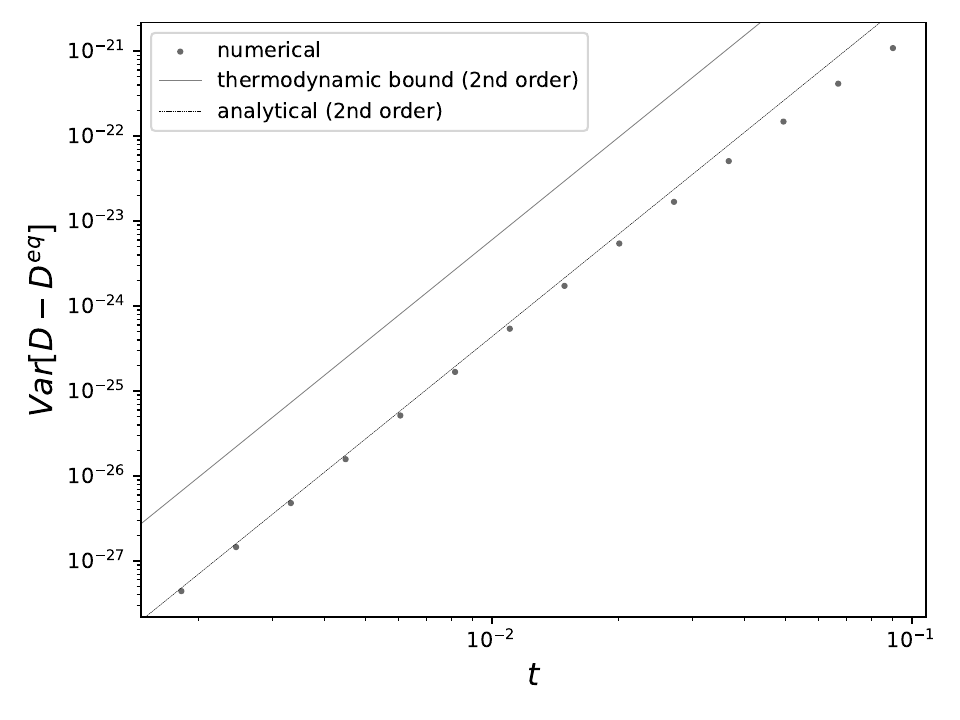}    \caption{\textbf{Quadratic example.} Variance of the nonequilibrium impact, here for the quadratic non-reciprocal model corresponding to the force field $\boldsymbol{F}=[-\beta x,\, \gamma (\mu_1 x^2 - \mu_2 -y)]$, with $\beta = 1$, $\gamma = 3$, $\mu_1=2/15$, $\mu_2 = 2$, and $T=5$. The perturbations parameters are here $\epsilon = 0.01$ and $k=0.01$.}
    \label{fig:A1}
\end{figure}

This model has been introduced in the main text, here is just the supplementary figure A\ref{fig:A1}.

\section{Random nonlinear force fields.}
The analytical results and thermodynamic bound introduced in this manuscript have also been numerically tested on a more general class of nonlinear models beyond the examples discussed in the main text. These are force fields in two dimensions $(x,y)$ of the form
\begin{equation}\label{random fields}
\begin{cases}
        F_x = f\,\left[ K_x(x,y)\, e^{- \left(\frac{x^2+y^2}{2 \lambda}\right)} +C \left( e^{-\beta x} -e^{\beta x} \right) \right],\\
        F_y = f\,\left[ K_y(x,y)\, e^{- \left(\frac{x^2+y^2}{2 \lambda}\right)} +C \left( e^{-\beta y} -e^{\beta y} \right) \right],
\end{cases}
\end{equation}
where $K(x,y)=k_{00}+k_{10}x+k_{01}y+\frac{1}{2}k_{20}x^2+k_{11}xy + ... + \frac{1}{6}k_{03}y^3 $ is a 3rd-order Taylor polynomial around the origin $(0,0)$, the parameter $\lambda$ controls the polynomial nonlinearity, $f$ its strength, and the term $C \left( e^{-\beta x} -e^{\beta x} \right)$ with constants $C>0$ and $\beta>0$ ensures the density to decay fast when moving far from the origin. Note that for each numerical experiment the coefficients of the polynomials $K_x$ and $K_y$ are randomly sampled independently. Two replicas are considered here below, both with confining parameters $C = 0.18$ and $\beta=3.5$, strength $f= 3$, and decay $\lambda=5$. A Laplacian smoothing at the border of the numerically considered square region of side length $l= 15$ has been performed to implement periodic boundary conditions, which is useful for the implementation of gradients with python numpy. The probability currents at the border introduced by such boundary conditions should be negligible.

\subsection*{Example 1}
The parameters here are 

\scriptsize
\begin{center}
$K_x$ :
\begin{tabular}{ c | c | c | c | c | c | c | c | c | c}
 $k_{00}$ & $k_{10}$ & $k_{01}$ & $k_{20}$ & $k_{02}$ & $k_{11}$ & $k_{30}$ & $k_{03}$ & $k_{21}$ & $k_{12}$\\
 \hline 
1.258455 & -0.33743 & -0.52390 & 0.794866 & 0.836348 & -0.57036 & 1.334669 & 0.466968 & -0.48637 & -1.24469   
\end{tabular}
\end{center}

\begin{center}
$K_y$ :
\begin{tabular}{ c | c | c | c | c | c | c | c | c | c}
 $k_{00}$ & $k_{10}$ & $k_{01}$ & $k_{20}$ & $k_{02}$ & $k_{11}$ & $k_{30}$ & $k_{03}$ & $k_{21}$ & $k_{12}$\\
 \hline 
-0.84271 & -0.06908 & -0.35831 & -1.44971 & 0.144042 & 1.096603 & -0.30986 & -0.42492 & -0.57323 & 0.752307   
\end{tabular}
\end{center}
\normalsize

\begin{figure}
    \centering
    \includegraphics[trim={1cm 0 2cm 0},clip,scale=0.65]{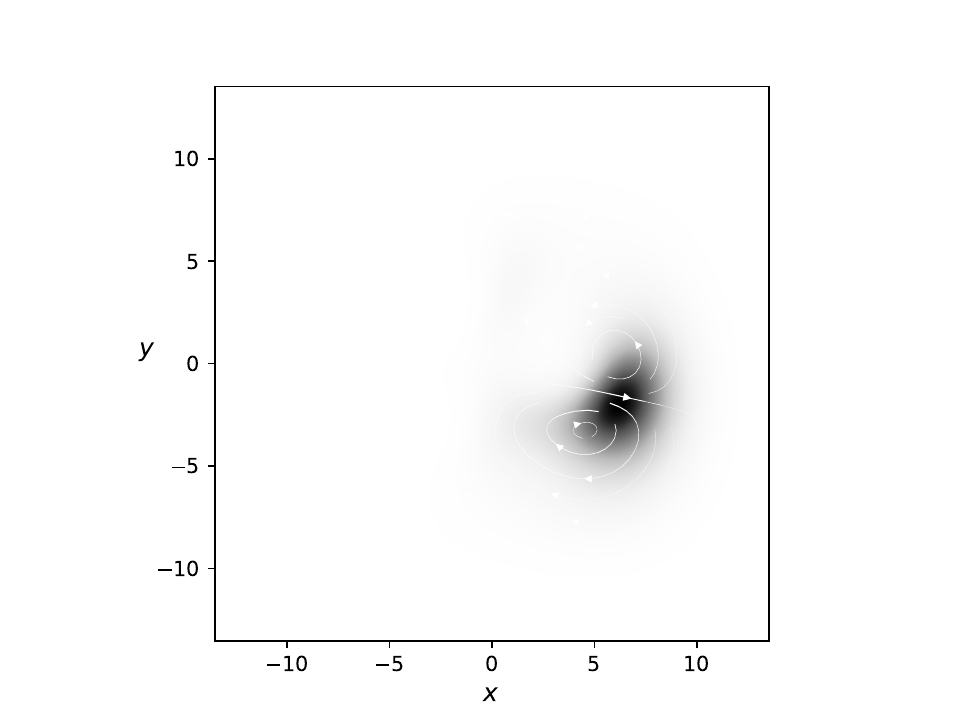} \includegraphics[trim={1cm 0 2cm 0},clip,scale=0.65]{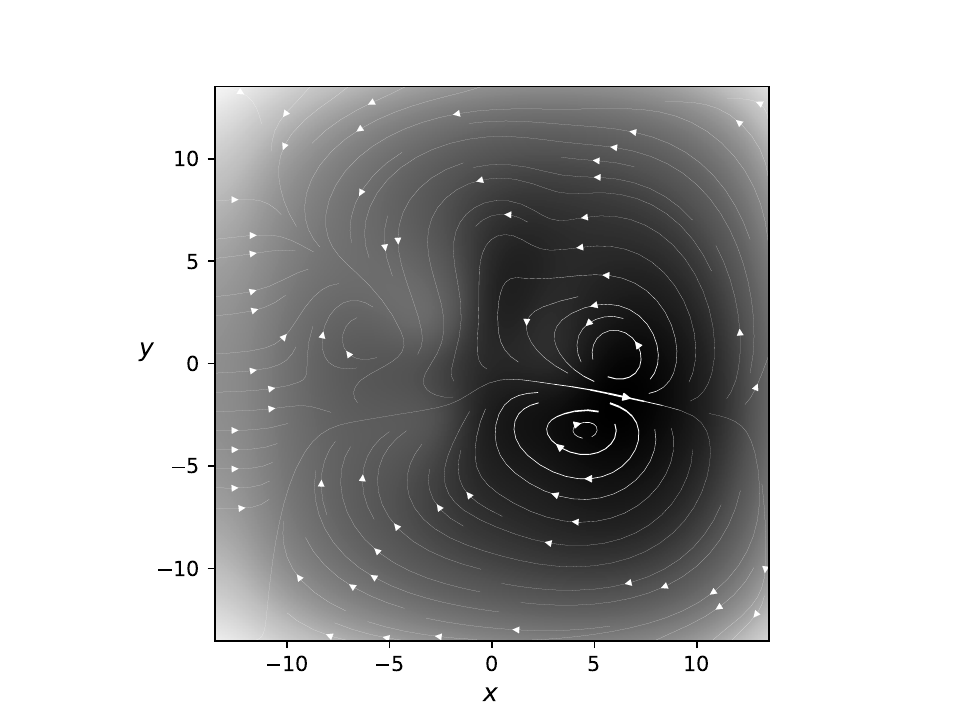}   \caption{\textbf{Random nonlinear forces. Example 1.} Plot of the steady-state density and nonequilibrium currents for the Example 1, with the density represented as colored background, normalized in linear (left) and log scale (right) to highlight the nonlinearities.}
    \label{fig:A2}
\end{figure}

\begin{figure}
    \centering
    \includegraphics[trim={0.4cm 0 0 0},clip,scale=0.55]{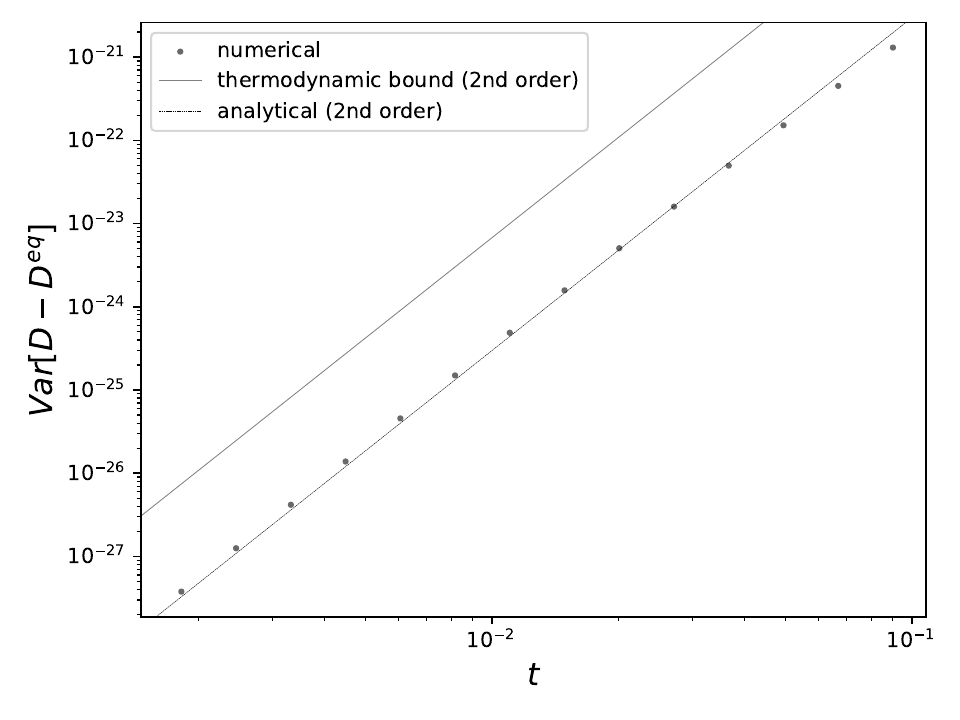}    \caption{\textbf{Thermodynamic bound on the random nonlinear example 1.} Thermodynamic bound and analytical result for the nonequilibrium impact on the divergence for this nonlinear example 1. The perturbations parameters are $\epsilon=0.01$ and $k = 0.01$. The small discrepancies observed likely arise from the discreteness of the estimation procedure, as indeed a smaller precision leads to wider gaps.}
    \label{fig:A3}
\end{figure}

\subsection*{Example 2}
The parameters here are 

\scriptsize
\begin{center}
$K_x$ :
\begin{tabular}{ c | c | c | c | c | c | c | c | c | c}
 $k_{00}$ & $k_{10}$ & $k_{01}$ & $k_{20}$ & $k_{02}$ & $k_{11}$ & $k_{30}$ & $k_{03}$ & $k_{21}$ & $k_{12}$\\
 \hline 
0.152015 & 0.596820 & 0.409914 & 0.481014 & -1.01156 & 1.648964 & -2.20127 & -0.57980 & -1.71535 & 0.422695  
\end{tabular}
\end{center}

\begin{center}
$K_y$ :
\begin{tabular}{ c | c | c | c | c | c | c | c | c | c}
 $k_{00}$ & $k_{10}$ & $k_{01}$ & $k_{20}$ & $k_{02}$ & $k_{11}$ & $k_{30}$ & $k_{03}$ & $k_{21}$ & $k_{12}$\\
 \hline 
-0.57089 & 0.585498 & 1.157988 & 0.420783 & -0.35388 & -0.54776 & 0.439099 & -1.68254 & 1.118676 & 2.662532   
\end{tabular}
\end{center}
\normalsize

\begin{figure}
    \centering
    \includegraphics[trim={1cm 0 2cm 0},clip,scale=0.65]{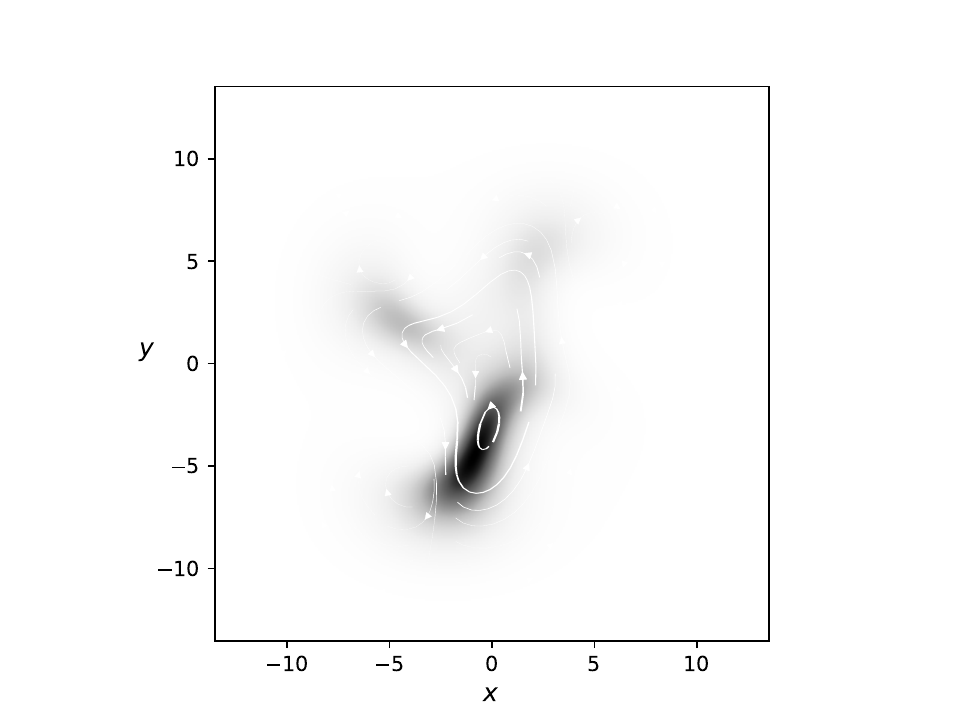} \includegraphics[trim={1cm 0 2cm 0},clip,scale=0.65]{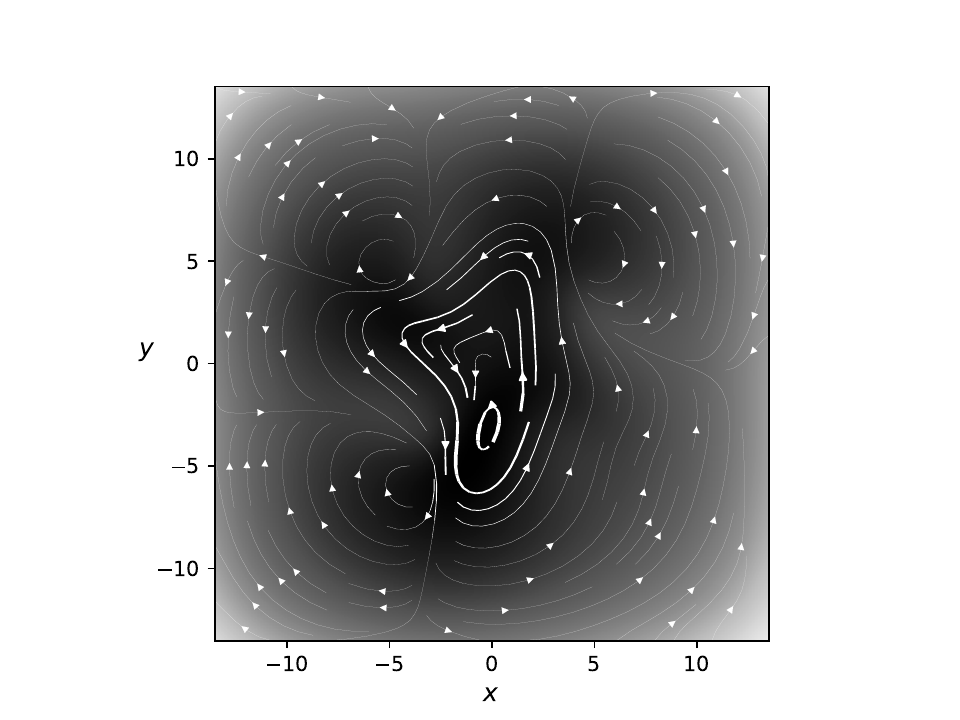}   \caption{\textbf{Random nonlinear forces. Example 2.} Plot of the steady-state density and nonequilibrium currents for the Example 2.}
    \label{fig:A4}
\end{figure}

\begin{figure}
    \centering
    \includegraphics[trim={0.4cm 0 0 0},clip,scale=0.55]{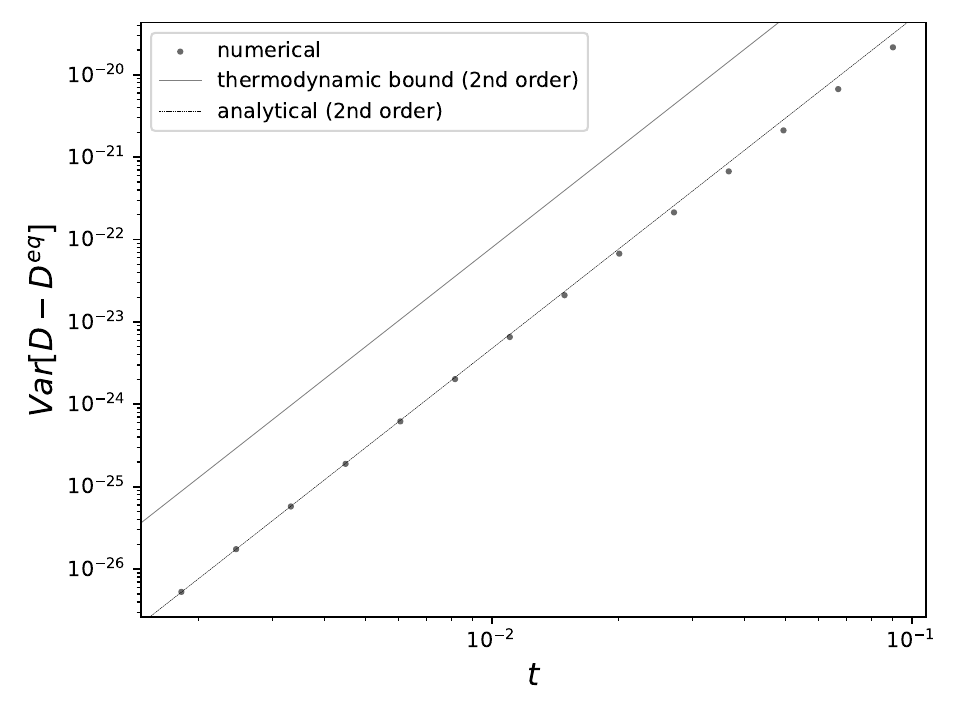}    \caption{\textbf{Thermodynamic bound on the random nonlinear example 2.} Thermodynamic bound and analytical result for the nonequilibrium impact on the divergence for this nonlinear example 2.}
    \label{fig:A5}
\end{figure}

\bibliography{apssamp}